\documentclass[
	aps, pra,
        superscriptaddress,
        twocolumn,
	10pt,
	floatfix, 
	floatfix,nobibnotes,
        nofootinbib,
	tightenlines
]{revtex4-2}
\usepackage[final]{graphicx}
\usepackage{times,bbm,amsmath,amssymb}
\usepackage{epsfig,color}
\usepackage{xcolor}
\usepackage{hyperref}
\hypersetup{
    colorlinks = true
}
\usepackage{cleveref}
\usepackage{microtype}

\usepackage{booktabs}
\usepackage{float}
\usepackage[caption = false]{subfig}

\usepackage[greek,english]{babel}
\usepackage{thumbpdf,enumerate}
\usepackage{booktabs}
\usepackage{sidecap}
\usepackage{pst-grad,bm}
\usepackage{epigraph}
\usepackage{longtable}
\usepackage{ulem}
\usepackage{acronym}
\usepackage{physics}
\usepackage{dsfont}
\usepackage{tabularx}
\usepackage{mathtools}
\usepackage{physics}

\usepackage{easyReview}
\usepackage{hyperref}
\usepackage{comment}

\begin{document}

\title{Optimal distillation of photonic indistinguishability} 

\author{Francesco Hoch}
\affiliation{Dipartimento di Fisica - Sapienza Universit\`{a} di Roma, P.le Aldo Moro 5, 00185 Roma, Italy}

\author{Anita Camillini}
\affiliation{CINECA Consorzio Interuniversitario, Via Magnanelli 6/3, 40033 Casalecchio di Reno, Italy}
\affiliation{International Iberian Nanotechnology Laboratory (INL)
 Av. Mestre José Veiga s/n, 4715-330 Braga, Portugal
}
\affiliation{Centro de F\'{i}sica, Universidade do Minho, Campus de Gualtar, 4710-057 Braga, Portugal}

\author{Giovanni Rodari}
\affiliation{Dipartimento di Fisica - Sapienza Universit\`{a} di Roma, P.le Aldo Moro 5, 00185 Roma, Italy}

\author{Eugenio Caruccio}
\affiliation{Dipartimento di Fisica - Sapienza Universit\`{a} di Roma, P.le Aldo Moro 5, 00185 Roma, Italy}

\author{Gonzalo Carvacho}
\affiliation{Dipartimento di Fisica - Sapienza Universit\`{a} di Roma, P.le Aldo Moro 5, 00185 Roma, Italy}

\author{Taira Giordani}
\affiliation{Dipartimento di Fisica - Sapienza Universit\`{a} di Roma, P.le Aldo Moro 5, 00185 Roma, Italy}

\author{Riccardo Albiero}
\affiliation{Istituto di Fotonica e Nanotecnologie, Consiglio Nazionale delle Ricerche (IFN-CNR), 
Piazza Leonardo da Vinci, 32, 20133 Milano, Italy}

\author{Niki Di Giano}
\affiliation{Dipartimento di Fisica, Politecnico di Milano, 
Piazza Leonardo da Vinci, 32, 20133 Milano, Italy}
\affiliation{Istituto di Fotonica e Nanotecnologie, Consiglio Nazionale delle Ricerche (IFN-CNR), 
Piazza Leonardo da Vinci, 32, 20133 Milano, Italy}

\author{Giacomo Corrielli}
\affiliation{Istituto di Fotonica e Nanotecnologie, Consiglio Nazionale delle Ricerche (IFN-CNR), 
Piazza Leonardo da Vinci, 32, 20133 Milano, Italy}

\author{Francesco Ceccarelli}
\affiliation{Istituto di Fotonica e Nanotecnologie, Consiglio Nazionale delle Ricerche (IFN-CNR), 
Piazza Leonardo da Vinci, 32, 20133 Milano, Italy}

\author{Roberto Osellame}
\affiliation{Istituto di Fotonica e Nanotecnologie, Consiglio Nazionale delle Ricerche (IFN-CNR), 
Piazza Leonardo da Vinci, 32, 20133 Milano, Italy}

\author{Marco Robbio}
\affiliation{International Iberian Nanotechnology Laboratory (INL)
 Av. Mestre José Veiga s/n, 4715-330 Braga, Portugal
}
\affiliation{Quantum Information and Communication, Ecole polytechnique de Bruxelles, CP 165/59, Université libre de Bruxelles (ULB),
1050 Brussels, Belgium}

\author{Leonardo Novo}
\affiliation{International Iberian Nanotechnology Laboratory (INL)
 Av. Mestre José Veiga s/n, 4715-330 Braga, Portugal
}
\affiliation{Centro de F\'{i}sica, Universidade do Minho, Campus de Gualtar, 4710-057 Braga, Portugal}

\author{Nicol\`o Spagnolo}
\affiliation{Dipartimento di Fisica - Sapienza Universit\`{a} di Roma, P.le Aldo Moro 5, 00185 Roma, Italy}

\author{Ernesto F. Galv\~ao}
\affiliation{International Iberian Nanotechnology Laboratory (INL)
 Av. Mestre José Veiga s/n, 4715-330 Braga, Portugal
}

\affiliation{Instituto de F\'isica, Universidade Federal Fluminense, Av. Gal. Milton Tavares de Souza s/n, Niter\'oi, RJ, 24210-340, Brazil}

\author{Fabio Sciarrino}
\affiliation{Dipartimento di Fisica - Sapienza Universit\`{a} di Roma, P.le Aldo Moro 5, 00185 Roma, Italy}

\begin{abstract}

Imperfect photons' indistinguishability limits the performance of photonic quantum communication and computation . Distillation protocols, inspired by entanglement purification, enhance photons' indistinguishability by leveraging quantum interference in linear optical circuits.  In this work, we present a three-photon distillation protocol optimized to achieve the maximum visibility gain, which requires consideration of multi-photon effects such as collective photonic phases. We employ interferometers with the minimum number of modes, optimizing also over the protocol's success probability. The developed protocol is experimentally validated with a platform featuring a demultiplexed quantum dot source interfaced with a programmable eight-mode laser-written integrated photonic processor. We achieve indistinguishability distillation with limited photonic resources and for several multi-photon distinguishability scenarios. This work helps to strengthen the role of distillation as a practical tool for photon-based quantum technologies.

\end{abstract}

\maketitle

\section*{Introduction}

\begin{figure*}[ht!]
    \centering
    \includegraphics[width=0.95\linewidth]{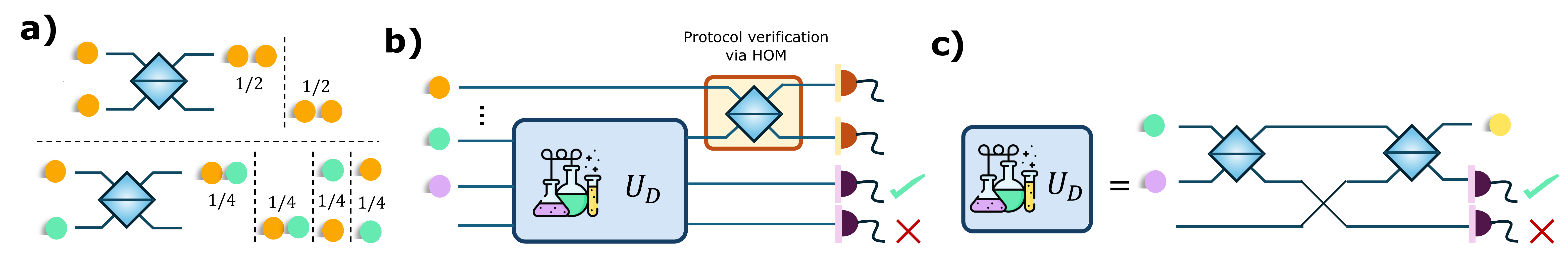}
    \caption{\textbf{Conceptual scheme of the indistinguishability distillation protocol.} a) Photon indistinguishability and the Hong-Ou-Mandel (HOM) effect. When two fully indistinguishable photons interfere on a balanced beamsplitter (BS), there is always photon bunching at the output. Conversely, if the photons are distinguishable, they can exit the BS in different ports.
    b) In a distillation protocol, the indistinguishability of input single-photon resource states is enhanced by having a subset of them interfere with auxiliary photons in a linear optical circuit. Photon-counting measurements in the auxiliary modes herald the generation of \textit{distilled} photons displaying higher mutual indistinguishability with the upper photons. This effect can be verified via a direct measurement of the HOM visibility. c) The optimal distillation circuit proposed in Eq.~(\ref{eq:mat_dist}), which takes as input two single-photon states with real-valued pairwise inner products associated with the HOM visibilities. Such a distillation unitary corresponds to the parameters $S = 1$ and $\varphi_u = 0$ and is optimal when the associated Gram matrix is parametrized by $V_{12} = V_{13} = V_{23}=V_\text{input}$ and $\varphi = 0$.
    }
    \label{fig:dist_gen_circ}
\end{figure*}

Photon indistinguishability is crucial to photon-based quantum computing and communication technologies \cite{Flamini2018}. Indeed, a high degree of indistinguishability is necessary for the correct implementation of quantum gates \cite{Knill2001, Kok_rev_2007, Obrien_rev, Bouchard2020, HOM_swap}, for the generation of entangled states employed in quantum computing \cite{Briegel2009, Bartolucci2023, Chen24, Cao24, Pont2024} and for enabling secure communications in quantum networks based on teleportation and entanglement swapping \cite{Gisin2007, Pirandola2015, Hu2023, Carvacho2017}. Additionally, photon indistinguishability underpins the computational complexity of Boson Sampling architectures \cite{AA_2010, Brod_rev}, and also of recently proposed models going beyond linear optics \cite{polacchi2025, Hoch2025}. In light of these applications, numerous strategies have been developed to harness this intrinsically quantum feature in photonic information processing \cite{Spagnolo2013, Menssen_17,Agne_17, Brod_19, Giordani2020, Pont_22, Seron2023, Rodari2025LAI}. 
However, suboptimal photonic indistinguishability can significantly hinder the performance of quantum photonic devices and limit their scalability \cite{Sund_24}, contributing to errors in the implementation of quantum protocols. As a result, the quantum photonics community has prioritized the development of sources capable of producing
highly indistinguishable photons \cite{Senellart2017, Caspani2017, Somaschi2016, Lohdal_QD} and, more generally, of platforms ensuring the preservation of photonic indistinguishability during information processing in quantum optical circuits.

Recently, photonic indistinguishability distillation protocols have emerged as a strategy to overcome the aforementioned challenges
associated with partial indistinguishability. 
This technique, inspired by methods for state and entanglement purification in qubit systems \cite{ Macchiavello_99, Pan2003, DeMartini_04, Kitaev_2005, Salart_20210}, involves the selective enhancement of indistinguishability using heralding measurements, starting from a set of partially distinguishable photons.
The core principle of these protocols is to leverage multi-photon interference in linear optical circuits, specifically designed for this task \cite{sparrow2018quantum, Marshall_22, Somhorst25, Saied25}, to probabilistically generate photonic states with enhanced indistinguishability.
In this context, resource-efficient strategies are fundamental for scaling up the technology while reducing error rates.
An experimental demonstration implementing the protocol proposed in Ref.~\cite{sparrow2018quantum} was recently reported in Ref.~\cite{Carosini_24}. The authors present an instance of a distillation procedure designed to improve the
indistinguishability between two photons, starting from four partially distinguishable ones.
An important goal is to devise distillation schemes for general distinguishability scenarios and capable of operating on a reduced number of initial photonic resource states, towards the application of such distillation techniques as robust tools for error mitigation in photonic 
quantum protocols.

This work presents a distillation protocol which, contrary to previous proposals, takes into account effects linked to the quantum interference properties of multi-photon states in the presence of partial distinguishability, such as non-trivial triad phases \cite{Menssen_17, Schesnovic19, Menssen_22}.
The method proposed here is optimal in the sense that, given an initial multi-photon resource state, it identifies the minimal multi-port optical circuit that maximizes the enhancement in indistinguishability.
Specifically, we identify a family of equivalent interferometers that maximize the possible gain for general initial distinguishability scenarios, and then further optimize by finding the interferometer design with the highest probability of success.
We test our optimal design in a three-photon experiment carried out employing a demultiplexed quantum dot source interfaced with a universal and programmable 8-mode integrated photonic processor. The initial multi-photon resource state is prepared in different distinguishability configurations by tuning both the polarization and the time-of-arrival degrees of freedom of each photon. In such a way, the performance of the distillation protocol, in terms of indistinguishability gain and success probability, was validated across several input conditions.
These results confirm the potential of distillation protocols as a promising tool to aid photon-based quantum information protocols in the presence of imperfect sources.

The manuscript is organised as follows. The first sections are dedicated to the theoretical concepts of the distillation protocol employed here. They include an introduction to the formalism of the paper, followed by a description of the procedure to identify the optimal distillation circuit for a given distinguishability scenario. Then, the experimental apparatus to implement and validate the protocol is described. Finally, in the third section, the experimental results showing the correct operation of the proposed distillation protocol are provided, followed by a summary of the obtained results and an overview of future prospects.

\section*{Theoretical background}

Photon distinguishability arises when photons differ in any of their internal degrees of freedom, such as polarization, frequency, or arrival time; even if they share the same spatial mode. In the simplest example of two photons entering different input ports of a balanced beam splitter, full indistinguishability yields perfect bunching, i.e. the photons always exit in the same port. This is known as the Hong-Ou-Mandel (HOM) effect \cite{Hong1987}.
If the photons are fully distinguishable, meaning that they have orthogonal internal states, they do not interfere, and, as classical particles would, exit opposite ports with probability $1/2$. Partially distinguishable photons lie in-between these two limits. The degree of two-photon interference is quantified by the visibility $V_{12}$, defined
as the reduction of coincidence counts relative to the classical case. For pure internal states $|\psi_1\rangle,|\psi_2\rangle$, one finds $V_{12}=|\langle\psi_1|\psi_2\rangle|^2$. Equivalently, the coincidence probability after a balanced beam splitter is $P_{c}= (1-V_{12})/2$. Thus $V_{12}=1$ (identical internal states) gives $P_{c} = 0$, whereas $V_{12}=0$ (orthogonal internal states) gives $P_{c}=1/2$, as shown in Fig.\ref{fig:dist_gen_circ}a.

These measurements extend to each pair in a multi-photon input: for $N>2$, one considers all pairwise visibilities $V_{ij}$. In a general linear interferometer, the unitary transformation acts only on the spatial modes and leaves internal states unchanged. The output probabilities are then obtained by summing over all ways of assigning input photons to output modes, with each term weighted by the (possibly complex) inner products of their internal states.

Formally, for an $N$-photon state with pure internal states, one can give a complete description of the multi-photon indistinguishability scenario using the Gram matrix formalism \cite{Oszmaniec2024}. A Gram matrix $\mathcal{G}$ encodes the complex pairwise photonic distinguishability, with matrix elements $\mathcal{G}_{i,j} = \braket{\psi_i}{\psi_j}$, where $\ket{\psi_i}$ is a pure spectral function describing the state of all the internal degrees of freedom of the $i$th photon \cite{Rodari2025LAI,rodari_24_counter}. 
The two extreme cases, i.e. the completely indistinguishable scenario and the distinguishable one, are respectively described by the Gram matrices $\mathcal{G}_{ij}=1$ and $\mathcal{G}_{ij}=\delta_{ij}$.

Partially distinguishable photons fit in-between these regimes, however, their outcome statistics is not a simple interpolation between the two aforementioned extreme cases \cite{Shchesnovich2015, Tichy2015}. 
The outcome probabilities depend not only on the pairwise visibilities $V_{ij}$, but also on non-trivial multiphoton effects associated with collective photonic phases, as we now discuss.

For the purpose of this work, let us consider the case of a three-photon experiment. By fixing the global phases of $|\psi_i\rangle$, one can always parameterize the off-diagonal elements in terms of their magnitudes, together with a single gauge-invariant phase \cite{Oszmaniec2024}. Concretely, by setting $\mathcal{G}_{12} = \sqrt{V_{12}}$, $\mathcal{G}_{23} = \sqrt{V_{23}}e^{i\varphi}$, and $\mathcal{G}_{13} = \sqrt{V_{13}}$, we may write
\begin{equation}
    \mathcal{G} = \begin{pmatrix}
    1 & \sqrt{V_{12}} & \sqrt{V_{13}}\\
    \sqrt{V_{12}} & 1 & \sqrt{V_{23}}e^{i\varphi}\\
    \sqrt{V_{13}} & \sqrt{V_{23}} e^{-i\varphi} & 1 
    \end{pmatrix}
    \label{eq:gram}
\end{equation}
Here, $\varphi = \mathrm{arg}(\mathcal{G}_{12}\mathcal{G}_{23}\mathcal{G}_{31})$ is the collective \textit{triad} phase. This is the phase of the unique nontrivial 3-photon Bargmann invariant $\Delta_{123} = \braket{\psi_1}{\psi_2}\braket{\psi_2}{\psi_3}\braket{\psi_3}{\psi_1}$ \cite{Oszmaniec2024}. This choice shows that the three-photon Gram matrix is fully specified by the three pairwise visibilities $\{V_{12},V_{23},V_{13}\} \in [0,1]$ and the single phase $\varphi \in [0,2\pi)$. The triad phase $\varphi$ influences the probabilities of three-photon experiment outcomes, describing collective interferometric effects that go beyond the predictions made using only the two-photon visibilities $V_{ij}$ \cite{Shchesnovich2015, Menssen_17}.

For mixed internal states $\{\rho_i\}$, one makes the analogous definition $V_{ij} = \mathrm{Tr}(\rho_i\rho_j)$, with $V_{ii}=1$ if $\rho_i$ is pure.
In this more general picture the Gram matrix is substituted by the three visibilities and the complex Bargmann invariant $\Delta_{123}=\text{Tr}(\rho_1 \rho_2 \rho_3)$. These quantities, together with the description of the interferometer, are sufficient to predict any outcome probability of a 3-photon linear-optical experiment with photon-counting measurements (see Appendix~\ref{app:mixed}). While the pairwise visibilities can be obtained via HOM tests, the Bargmann invariant can be reconstructed, for example, by postprocessing the data from a 3-mode Fourier interferometer \cite{Menssen_17, rodari_24_counter}. Knowledge of these quantities, which characterize the distinguishability of the input states, is crucial to the development of optimal distillation protocols proposed in this work.

\section*{Optimal distillation protocol}

\begin{figure*}[ht!]
\centering
\includegraphics[width=0.98\textwidth]{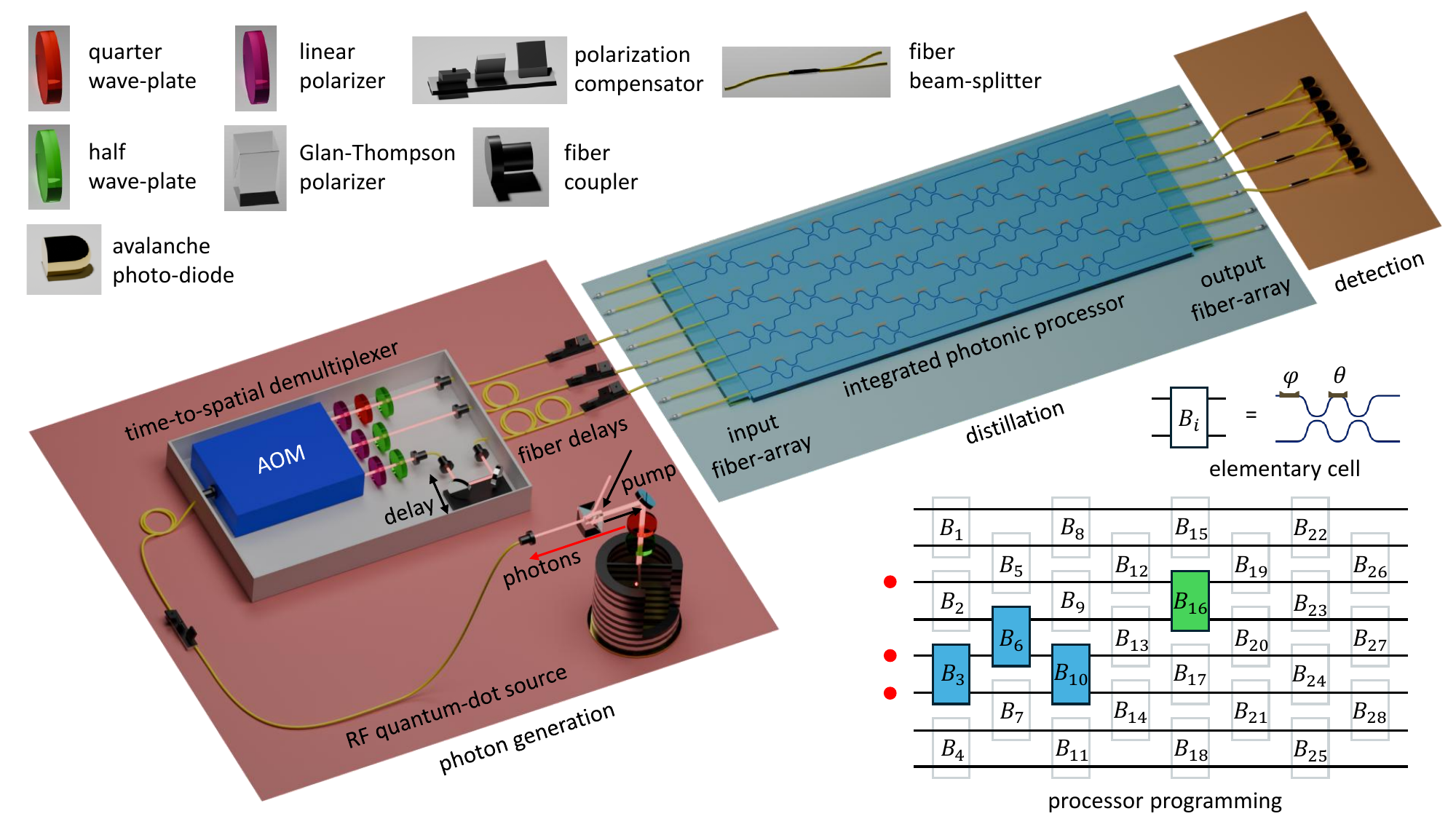}
\caption{\textbf{Experimental platform.} Scheme of the experimental setup employed for the implementation of optimal distillation protocols. A stream of single photons is generated via a quantum-dot source, pumped in the cross-polarization configuration via a pulsed laser, operating in the RF excitation regime. The stream of photons is converted into a multiphoton input state on multiple spatial modes via a time-to-spatial demultiplexer based on an acousto-optic modulator (AOM). Fiber delays are employed to temporally synchronize the input photons. Modulation of the multiphoton Gram matrix is performed in the polarization degree of freedom via polarization optics, and in the temporal degree of freedom via an adjustable delay line. The photons are then injected in the universal integrated photonic processor via an input fiber array and routed at the detection stage via an output one. Pseudo-photon number resolution up to $n=2$ is obtained probabilistically by adding a fiber beam splitter on each detected mode. Upper left inset: legend of the optical elements. Lower right inset: internal structure and programming of the photonic processor. Modes 3-6 are used to perform the distillation protocols. Elementary cells $B_{3}, B_{6}, B_{10}$ are used to implement the 3-mode transformations for the protocol, while $B_{16}$ is programmed to act as a beam splitter for the HOM characterization measurement. All other $B_{i}$ are programmed to act as the identity operation on the corresponding modes.}
\label{fig:setup}
\end{figure*}

The main intuition behind the inner workings of a distillation protocol can be traced back to the physics underlying the HOM effect \cite{Hong1987}. If two indistinguishable photons impinge onto the two input ports of a balanced beam splitter, they come out from the same output. Conversely, if the photons are distinguishable, then they exit with probability $p=1/2$ from different ports of the beam splitter (see Fig.~\ref{fig:dist_gen_circ}a).
In this two-photon scenario, the probability of detecting two photons in the same output is a measure of their mutual indistinguishability.
In a HOM experiment with two partially distinguishable photons, if we consider only events where the two photons came out from the same output, we expect to observe enhanced indistinguishability signatures in the two-photon state.

However, while this argument gives us an intuition as to why the distillation protocol may work, it does not apply to all regimes of partial distinguishability. 
The simplest non-trivial scenario involves three partially distinguishable photons, with the goal of distilling two photons with enhanced indistinguishability. 
Fig.~\ref{fig:dist_gen_circ}b depicts the general heralded circuit to implement a distillation protocol involving three photons, up to a permutation of the input photons.
There, two photons interfere within a linear-optical interferometer 
described by a unitary $U_D$ with elements $u_{ij}$, where $i$, ($j$) identify the input (output) modes of the linear optical transformation. All modes except the first are measured and a postselection is performed to herald events with one photon in the second mode and none in the others. The remaining photon, in the first mode, is then compared with the upper photon via HOM interference to characterise the indistinguishability gain of the distillation protocol.
In principle, one could consider distillation protocols where all three photons would undergo a non-trivial linear-optical interference with the aim of distilling two photons with higher indistinguishability by postselecting on observing one photon in one of the output modes. However, such a scenario can not be considered heralded, as it is unclear how to guarantee that the two distilled photons would occupy different spatial modes, since they would always have a probability of bunching together.

For a general linear-optical transformation described by unitary $U_D$, acting on an initial distinguishability scenario with pure input states described by Gram matrix $\mathcal{G}$, we analytically show in Appendix \ref{app:mixed}
that the final HOM visibility $V_f$
between the distilled photon and the upper photon, together with the success probability $P$ - i.e. the heralding probability - are given by: 
\begin{gather}
    \label{eq:Distilled_visibility}
    V_f = \frac{V_{12}S^2+V_{13}+2S\sqrt{V_{12}V_{13}V_{23}}\cos(\varphi+\varphi_u)}{1+S^2+2SV_{23}\cos(\varphi_u)}\\
    P = \abs{u_{12}u_{21}}^2\left(1+S^2+2SV_{23}\cos(\varphi_u)\right)
    \label{eq:probability}
\end{gather}
where $S = \abs{\frac{u_{11}u_{22}}{u_{21}u_{12}}}$ and $\varphi_u = \arg(u_{11}u_{22}u^*_{21}u^*_{12})$ are functions of the matrix elements of the unitary $U_D$ describing the interferometer. Note that a similar argument and derivation can also be given for the more general case of mixed states describing the internal degrees of freedom, as we briefly discuss in { Appendix~\ref{app:mixed}}. 
In what follows, in order to provide a quantifier for the improvement in terms of indistinguishability, one can define the gain $G$ of the distillation protocol as the difference between the final visibility and the maximum among the three initial visibilities
\begin{equation}
    G = V_f - \max(V_{12},V_{13},V_{23}).
    \label{eq:gain}
\end{equation}

Given a Gram matrix associated with the initial multi-photon indistinguishability scenario, the optimal distillation protocol will be associated with a unitary transformation $U_D$  maximizing the
gain $G$. {Up to a permutation of the input resources, an optimal unitary matrix achieving a gain $G \geq 0$ can always be found.} Such an \emph{optimal} unitary transformation can be identified via the following steps. First, one can perform a numerical maximisation of the distilled visibility as a function of the parameters $(S,\varphi_u) \in [0,\infty)\times [0,2\pi)$, considering these as free parameters. The correctness of this procedure is guaranteed by the unitary dilation theorem \cite{Levy2014}, which allows to embed an arbitrary $2\times2$ matrix as a submatrix of a unitary of dimension at most $4\times 4$, up to a suitable rescaling of the submatrix we want to embed.
Then, upon finding the optimal parameters $(S,\varphi_u)$, among all the unitary matrices corresponding to those parameters, a further optimization step can be performed so as to maximise the success probability of the protocol. In Appendix \ref{app:unitary} we present the analytical derivation of the optimal matrix $U_D$ starting from the parameters $(S,\varphi_u)$, and a proof of its optimality for indistinguishability distillation. It turns out that the resulting optimized unitary can be realized by a three-mode interferometer, instead of a four-mode interferometer as would be expected from the application of the unitary dilation theorem, as argued in Appendix \ref{app:unitary}.

Let us illustrate an optimal distillation protocol for the particular case of $V_{12} = V_{13} = V_{23}=V_\text{input}$ and $\varphi = 0$. For this case it can be shown that the optimal circuit is obtained for $S = 1$ and $\varphi_u = 0$. This choice of parameters corresponds to the unitary transformation:
\begin{equation}
    U_0 = \frac{1}{2}\begin{pmatrix}
       1 & 1 & \sqrt{2}\\
        1 & 1 & -\sqrt{2}\\
        \sqrt{2} & -\sqrt{2} & 0\\
    \end{pmatrix}.
    \label{eq:mat_dist}
\end{equation}
The corresponding interferometer is depicted in Fig.~\ref{fig:dist_gen_circ}c.  In this scenario, the optimal protocol is similar to that proposed in \cite{sparrow2018quantum}. However, this is not the case in more general scenarios with non-uniform visibilities or non-trivial collective photonic phases {(see Appendix~\ref{app:U0})}.
To show the importance of taking into account the complete information about partial distinguishability when designing distillation protocols, 
consider a similar scenario in which ${V_{12} = V_{13} = V_{23}=V_\text{input}}$ but with an associated triad phase $\varphi = \pi$, which can be realized if $V_\text{input}\leq 1/4$ \cite{Fernandes_Barg}. With the circuit represented by Eq. \eqref{eq:mat_dist}, now no gain in distillation is possible, i.e. $G\leq 0$. 
Indeed, one can show that the optimal parameters for the interferometer are now given by $S = 1$ and $\varphi_u = \pi$,  which can be implemented by switching the last two rows of the previous matrix. Practically, this is equivalent to post-selecting on the third mode instead of the second one.
In general, as argued in this section, for a given Gram matrix it is possible to perform a numerical maximisation of the distilled visibility to find the optimal parameters $S$, $\varphi_u$ and the optimal unitary matrix using the explicit formula derived in Appendix~\ref{app:unitary}. As we will see, this optimization can lead to significant improvements in the gain, when compared to the previously known protocol based on the unitary $U_0$ \cite{sparrow2018quantum, Carosini_24}. 

\section*{Experimental apparatus}

\begin{figure*}[ht]
    \centering
    \includegraphics[width=0.98\linewidth]{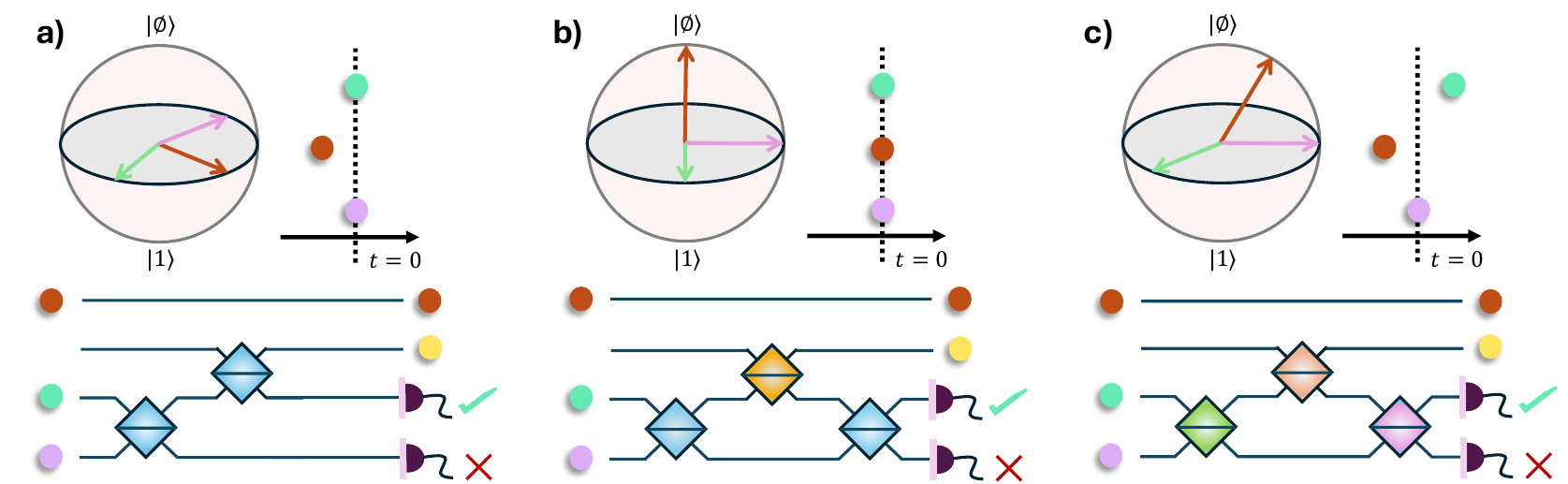}
    \caption{\textbf{Dependence of the optimal distillation circuit on the input indistinguishability scenario:} a) A real-valued Gram matrix preparation is associated with a simple transformation requiring only two optical elements. b) With a more general complex-valued Gram matrix preparation, associated to internal states that are Pauli eigenvectors, the optimal circuit is a balanced Fourier interferometer. c) For the  most general Gram matrix describing a three-photon preparation scenario using both polarization and time degrees of freedom, the optimal distillation circuit depends on the parameters $S$ and $\varphi_u$, and is found by optimizing  Eq.~\eqref{eq:Distilled_visibility} (See {Appendix \ref{app:unitary}}).}
    \label{fig:global_scheme}
\end{figure*}

The experimental implementation of the distillation protocol is performed using the QOLOSSUS photonic machine, first introduced in \cite{Rodari2025LAI} and used in subsequent works \cite{Hoch2025, rodari_24_counter}. Its overall setup is depicted in Fig.~\ref{fig:setup}, and it consists of three main stages: a photon generation and preparation stage, an evolution stage where the distillation protocol is implemented, and a detection stage with pseudo-photon-number-resolving detection.

A sequence of single-photon input states are generated by a commercial semiconductor quantum dot-based single-photon source (\textit{eDelight Quandela}) consisting of an InGaAs matrix located in a microscale electrically controlled micropillar cavity \cite{Gazzano2013,Thomas2021}. The source is excited in a resonance fluorescence (RF) configuration \cite{Somaschi2016} at a wavelength of $928.05\;\text{nm}$. The output photons are coupled to a single-mode fiber through a confocal microscope mounted on top of the cryostat and separated from the residual laser pump via a polarising beam splitter in a cross-polarisation configuration \cite{Somaschi2016}.
The train of single photons is then converted into the multi-photon input state through a bulk time-to-spatial demultiplexing setup (DMX), employing an RF-modulated acousto-optical modulator \cite{Pont2024}. The subsequent temporal synchronisation of the photon bunches is performed through a set of custom-length single-mode fibers.
After that, a generic three-photon indistinguishability scenario described by a Gram matrix $\mathcal{G}$ is prepared using both the polarisation and the time degree of freedom through a set of wave-plates and a time-delay line, which are suitably configured.

After the state preparation, the distillation protocol is carried out in an $8$-mode fully programmable polarisation-insensitive integrated photonic processor (IPP) \cite{Pentangelo2024}. As shown in Fig.~\ref{fig:setup}, the IPP is composed of a network of $28$ variable beam-splitters with arbitrary splitting ratios implemented using two cascaded balanced directional couplers and two thermo-optic phase shifters. The programmable phase shifters and the layout of the integrated processor provide us with universal controllability of the unitary transformation implemented with the linear-optical network of the IPP.
Specifically, for the distillation protocol, the variable beam splitters $3$, $6$ and $10$ are used to implement the optimal unitary matrix necessary for the distillation, and the beam splitter $16$ is programmed in a $50:50$ configuration to perform a HOM measurement to estimate the distilled visibility $V_f$, as shown in Fig.~\ref{fig:setup}.

Finally, detection of the photons is performed through a probabilistic pseudo photon-number-resolving setup of up to two photons in each output mode, implemented via additional in-fiber balanced beam splitters at each output, together with avalanche photodiodes (APD).

\section*{Experimental Results}

\begin{figure*}[ht!]
    \centering
    \includegraphics[width=1\linewidth]{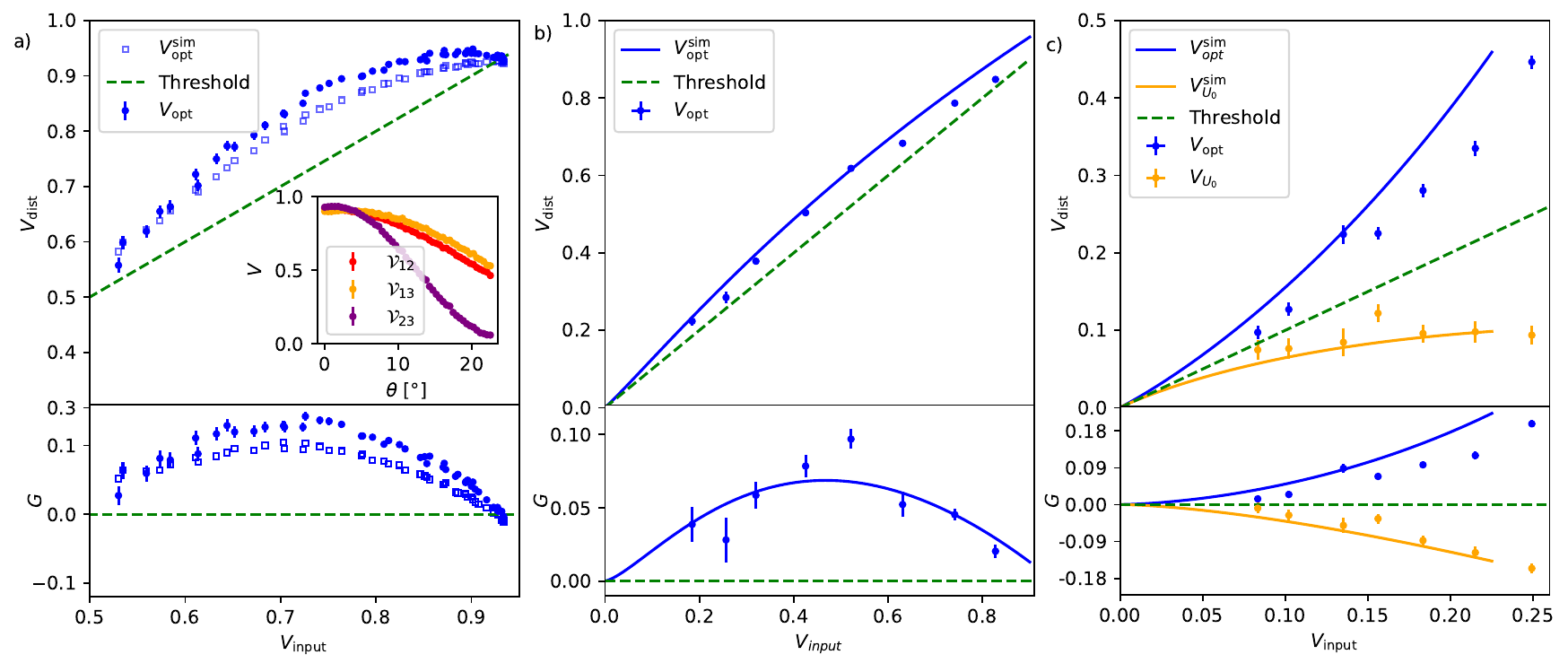}
    \caption{\textbf{Experimental characterization of the gain in distillation of indistinguishability.} Panel a) shows the distilled visibilities $V_{\mathrm{dist}}$ and the gain $G$ after the distillation process as a function of the maximum input pairwise visibility $V_{\mathrm{input}}$. The green dashed line describes the threshold above which a positive gain is obtained.  The input photons' polarization states are prepared as in Eq. \eqref{eq:polstates}. The filled blue circles (empty blue squares) represent the measured (simulated) results. The experiment simulation takes into account the measured Hong-Ou-Mandel (HOM) visibilities $V_{ij}$ between each pair of input polarization states, reported in the inset. The simulated visibilities are computed via a numerical model accounting for the intrinsic distinguishability of the QD source, the presence of a multi-photon component, and imperfect dialling of the unitary matrix. Panel b) shows the distilled visibilities $V_{\mathrm{dist}}$ and the distillation gain $G$ as a function of the mean input pairwise visibility $V_{\mathrm{input}}$ when the multi-photon state is prepared as in Eq. \eqref{eq:polstimetates}, where the time degree of freedom is used to obtain balanced pairwise input visibilities $V_{12} \sim V_{13} \sim V_{23} = V_{\mathrm{input}}$, and the angle $\theta$ is chosen in the interval $[0,\frac{\pi}{8}]$, for which the triad phase $\varphi=0$.
    Panel c) shows the distilled visibilities $V_{\mathrm{dist}}$ and the distillation gain $G$ as a function of the input pairwise visibility $V_{\mathrm{input}}$ when the multi-photon state is prepared as in Eq. \eqref{eq:polstimetates}, where the time degree of freedom is used to obtain balanced pairwise input visibilities $V_{12} \sim V_{13} \sim V_{23} = V_{\mathrm{input}}$, and the angle $\theta$ is chosen in the interval $[\frac{\pi}{8},\frac{\pi}{6}]$, for which the triad phase is $\varphi=\pi$.} 
    \label{fig:risultati_exp}
\end{figure*}

We experimentally validated the proposed three-photon distillation protocol for many different indistinguishability scenarios, obtained via fine control of the polarization state or time-of-arrival of the interfering photons. As said previously, each distinguishability scenario is described by a Gram matrix, and for each scenario an interferometer design with optimal gain was chosen.

\textbf{Polarisation preparation:} The first step is to test the feasibility of the distillation circuit, using the linear-optical unitary evolution $U_D=U_0$. 
We initially prepare all three photons in the $\ket{H}$ state. We then use half-wave plates to apply a polarization rotation to the second and third photons, whose polarisations are rotated respectively by an angle $\theta$ and $-\theta$, to obtain the polarisation states:
\begin{equation}
\begin{split}
    & \ket{\psi_1} = \ket{H} \\
    & \ket{\psi_2} = \cos(2\theta)\ket{H}+\sin(2\theta)\ket{V} \\
    & \ket{\psi_3} = \cos(2\theta)\ket{H}-\sin(2\theta)\ket{V}
    \label{eq:polstates}
\end{split}
\end{equation}
which result in the Gram matrix
\begin{equation}
    \mathcal{G} = \begin{pmatrix}
        1 & \cos(2\theta) &\cos(2\theta)\\
        \cos(2\theta) & 1 & \cos(4\theta) \\
        \cos(2\theta) & \cos(4\theta) & 1
    \end{pmatrix}.
\end{equation}

A null value for the 3-photon Bargmann invariant phase $\varphi$ results if we have $\theta \in [0,\frac{\pi}{8}]$. 
For this range of $\theta$ {and order of states}, the optimal gain is obtained for $S = 1$ and $\varphi_u = 0$, resulting indeed in the interferometer unitary $U_0$ of Eq.~\eqref{eq:mat_dist}.

The results of the protocol for the parameter $\theta$ varying from $0$ to $\pi/8$ are provided in Fig.~\ref{fig:risultati_exp}a.
In the inset, we show the experimentally measured pairwise HOM visibilities $V_{ij}$ between the input single-photon spectral functions $\{\ket{\psi_1}, \ket{\psi_2}, \ket{\psi_3}\}$ (orange, green and purple). In the main panel we show the final visibility $V_{opt}$ reached with the optimal unitary, as a function of the maximum input visibility $V_\text{input}$. 
Notably, we can see that for almost the whole range of the polarisation angular parameter $\theta$, the gain in visibility as defined in Eq. \eqref{eq:gain} is positive, reaching a maximum of $G_{max} = 0.1425 \pm 0.006 $ at $\theta=15.61 ^{\circ}$. Furthermore, the obtained visibility gains are compatible with a numerical simulation of the experiment carried out considering the typical sources of noise of our experimental apparatus, such as the presence of multi-photon components and imperfections in the implementation of the linear-optical circuit design.

\textbf{Polarisation and time preparation:}
To obtain a 3-photon indistinguishability scenario represented by a generic real-valued Gram matrix, i.e. any physical set of real-valued
inner products $\braket{\psi_i}{\psi_j}$, we need to control the polarisation and the time delay between one photon and the other two (See Fig.~\ref{fig:global_scheme}a). After adding a time-delay to the first photon, the three states after the preparation can be written as
\begin{equation}
\begin{split}
    & \ket{\psi_1} = \sqrt{(1-e^{-t/\tau})}\ket{H,0}+\sqrt{e^{-t/\tau}} \ket{H,t} \\
    & \ket{\psi_2} = \cos(2\theta)\ket{H,0}+\sin(2\theta)\ket{V,0} \\
    & \ket{\psi_3} = \cos(2\theta)\ket{H,0}-\sin(2\theta)\ket{V,0} 
\end{split}
\label{eq:polstimetates}
\end{equation}
The values of $\theta \in [0,\frac{\pi}{8}]$ and $\tau$ are controlled in such a way that the initial visibility is approximately equal for all three photon pairs $V_{ij} = V \in [0,1] \; \forall i\neq j$, and the 3-photon Bargmann invariant phase $\varphi = 0$.
As said previously, the optimal interferometer unitary matrix is the one given in Eq.~\ref{eq:mat_dist}.
The results of the distillation process are shown in Fig.~\ref{fig:risultati_exp}b.
As can be seen, there is a positive gain over the entire visibility range and the data is compatible with theoretical simulations.

\begin{figure*}[ht!]
    \centering
    \includegraphics[width=1\linewidth]{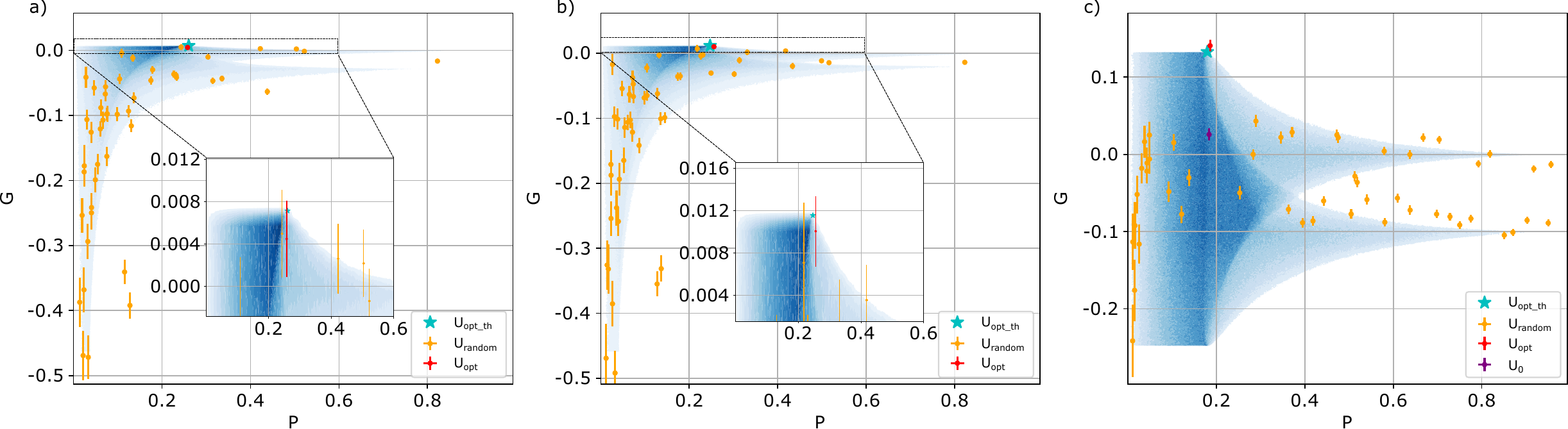}
    \caption{\textbf{Distilled HOM visibilities and success probabilities using randomly chosen interferometer designs, compared to the optimal design.}
     Visibility gain vs success probability for the optimal unitary $U_{\mathrm{opt}}$ (red dot) and $50$ random unitaries $U_{\mathrm{random}}$ (orange dots). In cyan the theoretical value for the optimal matrix.
     In blue, the theoretical distribution obtained by sampling from the random Haar distribution.
     \textbf{a)-b)} The photons are synchronised, and the polarisation states are prepared according to Eq. \eqref{eq:polstates} with $\theta\approx 0.61$ and $\theta\approx 0$ respectively (triad phase $\varphi=0$).
    \textbf{c)} The polarization state of the three injected photons is  $\vert L\rangle \vert V\rangle \vert A\rangle$ (triad phase $\varphi= -\pi/4$). In purple the visibility and the success probability for the unitary $U_0$, showing the importance of the knowledge of the triad phase for the correct optimization of the distillation.}
    \label{fig:risultati_exp2}
\end{figure*}

\textbf{Effect of the triad phase $\varphi$:}
Now, we analyze the most generic 3-photon indistinguishability scenario that corresponds to a Gram matrix with $\varphi \neq 0$.
To test how a non-zero triad phase affects the outcome of the distillation protocol, first, we use the same preparation strategy as in Eq. \eqref{eq:polstimetates}. When $\theta \in [\frac{\pi}{8},\frac{\pi}{6}]$, this corresponds to having $V_{ij} = V \in [0,0.25] \; \forall i\neq j$, while the 3-photon Bargmann invariant phase assumes the value $\varphi = \pi$.
The results are depicted in Fig.~\ref{fig:risultati_exp}c. As we can see in the orange curve, if we use the unitary matrix $U_0$ as in the previous experiments, the protocol fails as it is below threshold, i.e., it results in a negative gain $G$. If on the contrary we use the optimal unitary matrix associated to this scenario found by our approach -- i.e. the one given by $S = 1$ and $\varphi_u = \pi$ -- one can see that the circuit can recover a positive distillation gain (blue curve).

Another example of the triad phase effect is given by the Gram matrix where $V_{12} = V_{13} = V_{23} = 0.5$ and $\varphi = \pm \pi/4$ (see Fig.~\ref{fig:global_scheme}b). Such a preparation can be obtained using only the polarization degree of freedom of the three input photons, with polarization states given by $\ket{L/R}\ket{V}\ket{D}$. In this case, the distillation protocol implemented with the unitary transformation $U_0$ returns a distilled photon with no gain in visibility. If instead we use the optimal interferometer matrices, which have $S = 1$ and $\varphi_u = \mp 2\pi/3$ that corresponds to the three-dimensional Fourier matrix, we obtain a positive gain. The experimental results in this scenario are reported in Tab.~ \ref{tab: fase immaginaria}. As these two examples show, ignorance about the triad phase $\varphi$ can prevent us from choosing circuits with positive distillation gain.

\begin{table}[ht]
    \centering
    \begin{tabular}{|c|c|c c c|}
    \hline
        $\varphi$ & $\max(V_{\mathrm{input}})$ & $V_{U_0}$ & $V^{\text{tritter}}_{+}$& $V^{\text{tritter}}_{-}$ \\
        \hline
        $\pi/4$ & $0.469 (3)$ & $0.465 (7)$ & $0.344 (6)$& $0.618 (6)$\\
        $-\pi/4$ & $0.405 (2)$ & $ 0.451 (8)$ & $0.676 (8)$& $0.234 (6)$\\
        \hline
    \end{tabular}
    \caption{\textbf{Distilled HOM visibilities for complex-valued triad phases.} The polarization states of the injected three photons are $\vert L\rangle \vert V\rangle \vert D\rangle$ (triad phase  $\varphi= \pi/4$) and  $\vert L\rangle \vert V\rangle \vert A\rangle$ (triad phase  $\varphi= -\pi/4$). The second column shows the maximum input visibilities $\max(V_{\mathrm{input}})$ among the three possible photon pairs. The third column shows the visibilities $V_{U_0}$, obtained using the distillation unitary $U_0$ of Eq. \eqref{eq:mat_dist} that do not improve the visibilities of the input photons. The fourth  and fifth column shows the visibilities $V^{\text{tritter}}_{\pm}$, obtained with the unitary of Eq.~\eqref{eq:optimal_unitary} with $S=1$ and $\varphi_u = \mp 2\pi/3$. As we can see the optimal unitary for this Gram matrix preparation scenario is the one with the opposite sign of the triad phase of the preparation.}
    \label{tab: fase immaginaria}
\end{table}

\textbf{Optimality of the distillation matrix:} Finally, we test the optimality of the unitary transformation presented in { Appendix \ref{app:unitary}}, associated with a given Gram matrix preparation via the parameters $(S,\varphi_u)$. 
To do so, we compare the optimal gain $G$ and optimal success probability with those obtained from $50$ different randomly drawn $3\times3$ unitary matrices, all implemented in the IPP. The results associated to three specific Gram matrix preparations are depicted in Fig.~\ref{fig:risultati_exp2}. There, we show the behaviour of the distillation gain as a function of the protocol's success probability. The red dot represents the optimal distillation protocol, i.e. where the unitary transformation results in the maximum gain and the highest success probability. Instead, the orange points are associated to circuit designs corresponding to random matrices. In the blue shaded area, we represent the theoretical distribution of the visibility gain vs the success probability, obtained by numerically sampling  $10000$ unitary evolutions from the random Haar distribution for a fixed Gram matrix preparation. In cyan we depict the optimal point obtained numerically with the unitary $U_\textbf{opt}$ of Eq.~(\ref{eq:optimal_unitary}).

For all configurations under test, we note that the optimal unitary performs better than all other random unitary matrices, even though it is not the only unitary transformation resulting in a positive gain.
Furthermore, it can be seen that there are unitaries that have a higher success probability than the one we consider optimal, but this is explained by the fact that optimization on the success probability is only done after finding the circuit design having optimal gain in visibility. 

\section*{Discussion}

In this work, we obtain and experimentally implement a protocol tailored for the distillation of photon indistinguishability. The protocol is based on the interference of three
photons in a linear interferometer, where post-selection 
is performed on the output, mediating the increase in the indistinguishability among the unmeasured photons. 
In contrast to previous proposals, this work proposes a provably optimal protocol for a given multi-photon indistinguishability scenario. 
Optimality here is understood in the sense that the protocol achieves the highest possible visibility amongst the output pair of photons, {with respect to a well-defined association between input states and optical modes}.
A further optimization step is also performed among the unitaries that achieve the maximum gain in visibility in order to maximize the protocol's success probability. We note that this last optimization step can be done analytically, in a way that also minimizes the number of ancillary modes. {In this way, the protocol analyzed here is always able to provide a gain $G \geq 0$ with a maximized success probability 
up to the inclusion of an input state permutation.}

The proposed protocol is experimentally tested in a polarisation-independent integrated photonic processor using multi-photon states generated by a demultiplexed quantum dot source, in various distinguishability regimes implemented via manipulation of both polarization and time-of-arrival. We show the practical feasibility of implementing a distillation protocol within a typical photonic architecture, thus highlighting the possibility of integrating it into more complex protocols of photon-based quantum computation. In particular, the protocol is tested using different degrees of freedom to address different regimes of partial distinguishability of the photons, showing the dependence between photon preparation and the indistinguishability gain reached via distillation. The results show that the proposed unitaries are indeed optimal and demonstrate that for an optimal application of the protocol, the knowledge of the Gram matrix characterising the partial distinguishability of photons is essential. In particular, we show that the optimal protocol depends on the value of the collective photonic phase $\varphi$, a non-trivial multi-photon effect whose importance is often neglected in the literature. 
{One could argue that in practical near-term scenarios partially distinguishable photonic resources can be engineered in order to nullify the collective photonic phase - e.g. via accurate polarization control and filtering. However, as shown in Appendix \ref{app:U0}, we found numerical evidence that, even for real-valued Gram matrices, the optimization of the interferometer used for distillation leads to significantly better performance when compared to previously known protocols based on the matrix $U_0$ of Eq.~\eqref{eq:mat_dist}. This clearly shows the need for optimized distillation protocols even when triad phases do not play a role. 
Moreover, note that in practical scenarios a collective photonic phase could emerge due to imperfect polarization control, or due to effects related to other degrees of freedom, e.g. in distributed scenarios where sources with different spectral or temporal properties are interfaced.}

Our findings provide a novel perspective for the development of resource-efficient tools to achieve indistinguishability distillation within practical photonic platforms. Furthermore, we foresee that similar arguments could be extended without loss of generality to scenarios featuring a higher number of single-photon resource states. Nonetheless, the modest number of resources, in terms of additional optical modes and photon states, can make it feasible to integrate distillation routines within more complex photon-based protocols, thus opening up new approaches to error mitigation for quantum information processing in photonic platforms. 

\section*{Acknowledgements}
This work was supported by the European Union’s Horizon Europe research and innovation program under EPIQUE Project (Grant Agreement No. 101135288) and by ICSC – Centro Nazionale di Ricerca in High Performance Computing, Big Data and Quantum Computing, funded by European Union – NextGenerationEU; and by FCT – Fundaç\~{a}o para a Ciência e a Tecnologia (Portugal) via project CEECINST/00062/2018. M.R. is a FRIA grantee of the Fonds de la Recherche Scientifique – FNRS. L.N. acknowledges the financial support of the project
with the reference n.º 2023.15565.PEX, funded by national funds
through FCT – Fundação para a Ciência e a Tecnologia, I.P.
https://doi.org/10.54499/2023.15565.PEX.

\section*{Data Availability}

Data supporting the current study are available from the corresponding author on reasonable request.

\section*{Competing interests}

G.Co., F.C., and R.O. are co-founders of Ephos. The other authors declare no other competing interests.

\section*{Author Contributions}

F.H., A.C., G.R., T.G., M.R., L.N., N.S., E.F.G. and F.S. conceived the concept and experiment. 
R.A., N.D.G., F.C., G.Co., and R.O. fabricated the photonic chip and characterized the integrated device using classical optics. F.H., G.R., E.C., G.Ca., T.G.,  N.S. and F.S. carried out the quantum experiments and performed the data analysis. 
All the authors discussed the results and contributed to the writing of the paper.

\appendix

\section{Distilled visibility and extension to mixed states}
\label{app:mixed}

In this section, we initially derive the final visibility of the non-optimised protocol in Fig.\ref{fig:dist_gen_circ}b, following the theory introduced in \cite{Shchesnovich2015, Tichy2015}. We then extend it to the case of mixed internal photonic degrees of freedom.

Let us denote by $\rho_2,\rho_3$ the two photons impinging upon the three-mode linear interferometer $\hat{U}_D$, and by $\rho_d$ the distilled photon emerging from the interferometer through the first output mode - the successful detection pattern being $[1,1,0]$. Let $\rho_1$ be the photon used for the verification of the distillation effect. The HOM visibility between $\rho_1$ and the distilled photon $\rho_d$ is given by $\mathrm{Tr}(\rho_1 \rho_d)$, and it depends on the bunching events at the output of the two upper modes. Since $\rho_1$ and $\rho_d$ go through a balanced beam splitter - coupling only the spatial modes - the probability of observing both photons in the same mode - namely, bunching events - depends on the overlap between the two states
\begin{equation}
    p_{(1,1)\mapsto(2,0)} = p_{(1,1)\mapsto(0,2)} = \dfrac{1+\mathrm{Tr}(\rho_1 \rho_d)}{4}.
\end{equation}
Hence the full bunching probability can be related directly to the visibility, meaning that
\begin{equation}
    p_{bunch}=\frac{1+\operatorname{Tr}(\rho_1 \rho_d)}{2}\implies \operatorname{Tr}(\rho_1 \rho_d)=2p_{bunch}-1
\end{equation}
and such probability can be computed according to
\begin{equation}\label{eq.pBu}
    p_{bunch}=\frac{p_{(2,0,1,0)}+p_{(0,2,1,0)}}{p_{(1,1,1,0)}+p_{(2,0,1,0)}+p_{(0,2,1,0)}}, 
\end{equation}
where we recall that we perform a post-selection on 3-photon events with vacuum on the last mode.

In the ideal scenario where bosons are fully indistinguishable, the outcome probabilities are computed via matrix permanents. More precisely, assuming the linear interferometer exploited in this protocol is described by the unitary matrix $U\in\mathbb{C}^{m\times m}$, we can write the transition probabilities from an input state of $n$ indistinguishable particles described by a mode occupation vector $\boldsymbol{j}$ to an output state described by a mode occupation vector $\boldsymbol{k}$ by calculating the permanent
\begin{equation}
    p_{\boldsymbol{j}\mapsto \boldsymbol{k}}=\frac{1}{\boldsymbol{j}!\boldsymbol{k}!}|\operatorname{perm}(U_{\boldsymbol{j},\boldsymbol{k}})|^{2},
\end{equation}
where $[U]_{\boldsymbol{j},\boldsymbol{k}}$ is known as the scattering matrix. This matrix is made by the rows and columns of $U$ repeated according to the multiplicity of the input occupation vector and output occupation vector. Moreover, we adopt the notation $\boldsymbol{k}!=k_{1}!...k_{m}!$. 

For partially distinguishable single-photon pure input states in the first $n$ modes, corresponding to  a Gram matrix $\mathcal{G}$ , the scattering matrix has the form $M_{\boldsymbol{k}}=U_{(1,...,1,0,...),\boldsymbol{k}}$. Hence, the transition probabilities can be written as a tensor permanent \cite{Tichy2015}
\begin{equation}\label{eq: equation for probabilities}
    p_{\boldsymbol{j}\mapsto \boldsymbol{k}}=\frac{1}{\boldsymbol{k}!}\sum_{\sigma\in S_{n}}\prod_{i=1}^{n}\mathcal{G}_{i,\sigma(i)}\operatorname{perm}(M_{\boldsymbol{k}}\odot\bar{M}_{\boldsymbol{k}}^{(\sigma)}),
\end{equation}
where $S_{n}$ indicates the symmetric group of dimension $n$, and $\sigma$ represent a generic permutation. In addition, the notation $\odot$ represents the Hadamard product (also known as element-wise product) and $M_{\boldsymbol{k}}^{(\sigma)}$ is the matrix $M_{\boldsymbol{k}}$ where the columns where permuted according to $\sigma$.

We can therefore compute the probabilities in Eq.\ref{eq.pBu} in the more general scenario of partially distinguishable input photons via Eq.~ \ref{eq: equation for probabilities}. This yields the final visibility between states $\rho_1$ and $\rho_d$
\begin{equation}\label{eq.vf}
    V_f = \frac{V_{12}S^2+V_{13}+2S\sqrt{V_{12}V_{13}V_{23}}\cos(\varphi+\varphi_u)}{1+S^2+2SV_{23}\cos(\varphi_u)}
\end{equation}
where $S = \abs{\frac{u_{11}u_{22}}{u_{21}u_{12}}}$ and $\varphi_u = \arg(u_{11}u_{22}u^*_{21}u^*_{12})$ are functions of the matrix elements of the unitary $U_{D}$ describing the interferometer, $V_{ij} = |\langle\psi_i|\psi_j\rangle|^2$ are pairwise overlaps between the three pure states $\rho_{i}=|\psi_{i}\rangle\langle\psi_{i}|$, and $\varphi$ is the triad phase defined below Eq.~\eqref{eq:gram}.

When instead the spectral functions describing all the internal degrees of freedom of the photons are mixed, the Gram matrix formalism is in general not valid. In this case, we define the photon visibility and the 3-photon Bargmann invariant as
\begin{equation}
    V_{i,j} = \Tr(\rho_i \rho_j) \quad \Delta_{123} = \Tr(\rho_1\rho_2\rho_3)
\end{equation}
and the triad phase as before, $\varphi = \arg(\Delta_{123})$.
Notice, that for pure states, these definitions reduce to the ones introduced above in the paragraph. The main difference between pure and mixed states is that the modulus of the 3-photon Bargmann invariant is no longer deducible from the photon visibilities alone. Indeed, the relation is $\abs{\Delta_{123}}^2\leq V_{12}V_{13}V_{23}$, with equality holding only for pure states.
For mixed states, Eq.\ref{eq.vf} becomes
\begin{equation}
    V_f = \frac{V_{12}S^2+V_{13}+2S\abs{\Delta_{123}}\cos(\varphi+\varphi_u)}{1+S^2+2SV_{23}\cos(\varphi_u)}.
\end{equation}
The success probability of the protocol is given by the probability of observing the pattern $(\cdot, \cdot, 1,0)$, i.e. one photon in the third mode and vacuum on the last. This results in
\begin{align}
    P &= p_{(1,1,1,0)}+p_{(2,0,1,0)}+p_{(0,2,1,0)}\nonumber\\
    &=\abs{u_{12}u_{21}}^2\left(1+S^2+2SV_{23}\cos(\varphi_u)\right).  
\end{align}
The optimization procedure for the distillation protocol with mixed-state inputs is the same as the one presented in the main text for pure-state inputs.

\section{Optimal unitary transformation} \label{app:unitary}
In this section, we provide an algorithm that, for a given pair of parameters $S$ and $\varphi_u$, returns an associated unitary matrix that also maximises the protocol's success probability. Hence, after obtaining these parameters by maximizing the output visibility $V_f$, we can build the unitary $U_D$ as follows.

First, we want to prove that for any given $S$ and $\varphi_u$ a unitary transformation always exists using the unitary dilation theorem \cite{Levy2014}. For our purposes, the theorem can be expressed as:\\
\textbf{Unitary dilation theorem:} For any contraction matrix $A \in \mathbb{C}^{N \times N}$ (i.e. $\norm{A}_2\leq1$) there exists an unitary matrix $U\in \text{U}(2N)$ that contains $A$ as a submatrix. That matrix is called a dilation of $A$.
Moreover all the dilations $U$ are equivalent \footnote{In this context, the equivalence relation is defined as: $U\sim V$ if there exist two unitary matrices $U_1, U_2 \in U(N)$ such that $V = (\mathds{I}\oplus U_1)U(\mathds{I}\oplus U_2)$} to the matrix
\begin{equation}
    U = \begin{pmatrix}
        A & \sqrt{\mathds{I} - AA^\dagger}\\
        \sqrt{\mathds{I} - A^\dagger A} & -A^\dagger
    \end{pmatrix}.
    \label{eq:dilation}
\end{equation}
Given the parameters $(S,\varphi_u)$, we can define without loss of generality a matrix $M$ as
\begin{equation}
    M = \begin{pmatrix}
        1 & a\\
        b & ab Se^{i\varphi_u}
    \end{pmatrix}, 
\end{equation}

with $a,b\geq0$.
To apply the previous theorem, we need a contraction matrix so we can define 
\begin{gather}
    A = \frac{M}{c\norm{M}_2},\\
    \norm{M}_2 = {\scriptstyle\sqrt{\frac{1+a^2+b^2+a^2b^2S^2+\sqrt{1+a^2+b^2+a^2b^2S^2-4a^2b^2(1+S^2-2S\cos(\varphi_u))}}{2}}, }
\end{gather}
with $c\geq1$.

The matrix $A$ defined above is the most general contraction matrix with the required parameters $S$ and $\varphi_u$ so Eq.~(\ref{eq:dilation}) can be used to construct the unitary matrix $U$ for the distillation protocol.

To choose the free parameters $a,b,c$, we can find the one that maximises the success probability of Eq.~\eqref{eq:probability}, which is obtained for $a = b = \frac{1}{\sqrt{S}}$ and $ c = 1$.
With those parameters the final matrix is
\begin{equation}
    A = \frac{1}{\sqrt{1+S+\sqrt{2S\left(1+\cos(\varphi_u)\right)}}}\begin{pmatrix}
        \sqrt{S}& 1\\1 & \sqrt{S}e^{i\varphi_u}
    \end{pmatrix}
\end{equation}

Since $\norm{A}_2 = 1$ then the matrix $\sqrt{\mathds{I} - AA^\dagger}$ is singular and so the unitary matrix is equivalent to one in the form
\begin{equation}
    U = \begin{pmatrix}
        {A} & \begin{matrix}
             u_{13} & 0\\
             u_{23} & 0 \\
        \end{matrix}\\
        \begin{matrix}
            u_{31} & u_{32}\\
            0 & 0
        \end{matrix} & \begin{matrix}
            u_{33} & 0\\
            0 & 1
        \end{matrix}
    \end{pmatrix}
\end{equation}

This implies that we can discard a dimension and find a $3\times3$ dilation of the matrix $A$. 
The explicit form of the unitary matrix is:
\begin{widetext}
\begin{equation}
    U = \frac{1}{\sqrt{1+S+\sqrt{2S\left(1+\cos(\varphi_u)\right)}}}
    \begin{pmatrix}
        \sqrt{S}& 1 & \sqrt[4]{2S(1+\cos(\varphi_u))}\\[1.3ex]
        1 & \sqrt{S}e^{i\varphi_u} & - \sqrt[4] {\frac{S}{2(1+\cos(\varphi_u))}}\left(1+e^{i\varphi_u}\right)\\[1.3ex]
        \sqrt[4]{2S(1+\cos(\varphi_u))} & - \sqrt[4] {\frac{S}{2(1+\cos(\varphi_u))}}\left(1+e^{i\varphi_u}\right) & \frac{1+e^{i\varphi_u}-\sqrt{2S(1+\cos(\varphi_u))}}{\sqrt{2(1+\cos(\varphi_u))}}
    \end{pmatrix}
    \label{eq:optimal_unitary}
\end{equation}
\end{widetext}
and the corresponding success probability is 
\begin{equation}
    P = \frac{\left(1+S^2+2SV_{23}\cos(\varphi_u)\right)}{\left(1+S+\sqrt{2S\left(1+\cos(\varphi_u)\right)}\right)^2}
\end{equation}

\begin{figure}[ht]
    \centering
    \includegraphics[width=\linewidth]{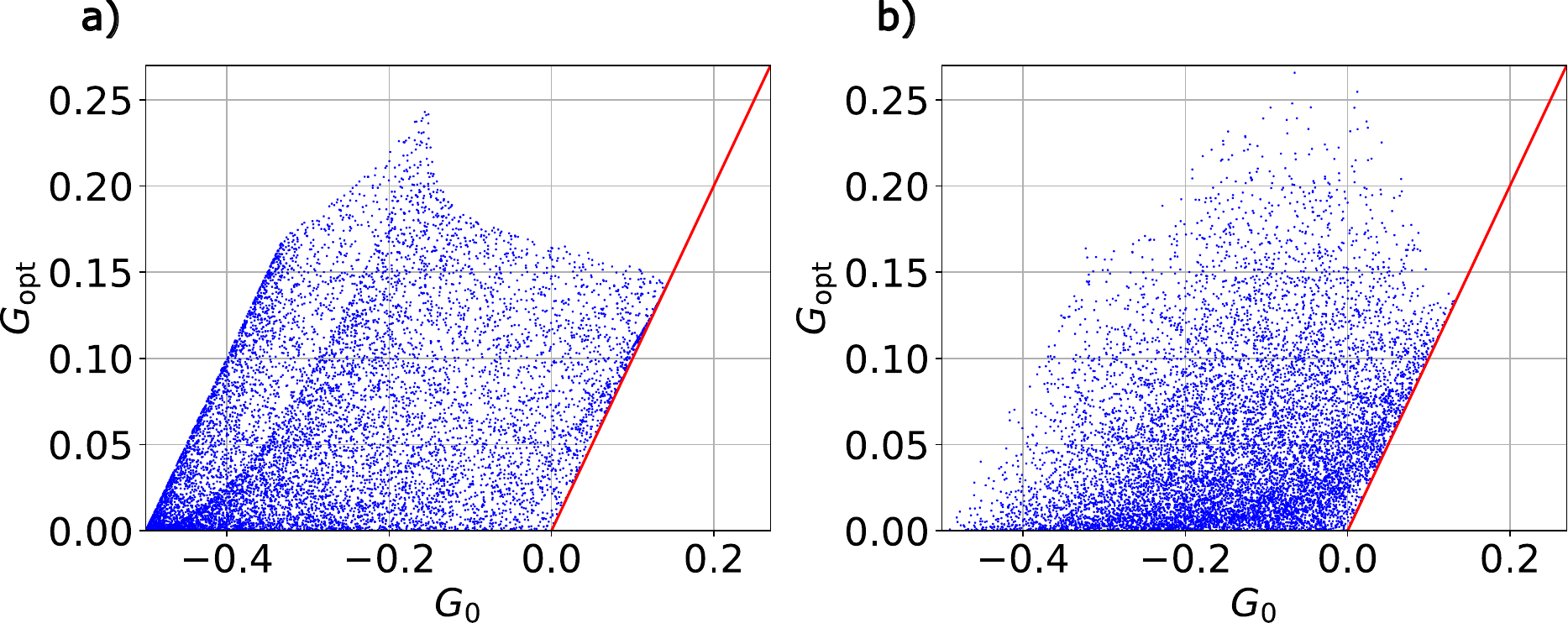}
    \caption{\textbf{Comparison of the gain using the matrix $U_0$ and the optimal one}. Optimal Gain $G_{opt}$ compared to that of the matrix $U_0$ for real Gram matrices \textbf{a)} and for complex matrices \textbf{b)}. In red, the line $G_{opt} = G_0$. The gain can almost always be improved compared to that of the matrix $U_0$ ( all the points are above the red line), showing that the knowledge of the Gram matrix is not only useful but, in most cases, essential to perform the distillation.}
    \label{fig:U0}
\end{figure}

{
\section{Non optimality of the $U_0$ matrix}
\label{app:U0}
In this section, we study the behaviour of the distillation protocol involving $U_{0}$, similar to that already proposed in \cite{sparrow2018quantum},  outside of the simplified model of equal visibilities. We prove that for distinguishability scenarios more complex than the naive model, the protocol is suboptimal or, in the worst case, leads to a negative gain, according to the definition of Eq. \eqref{eq:gain}. Figure~\ref{fig:U0} depicts the optimal gain $G_{opt}$ which can be obtained by employing the optimal distillation matrix found with our protocol, as a function of the gain obtained with the matrix $U_0$ ($G_0$), for a set of $10000$ randomly uniformly drawn real Gram matrices (Panel a) and complex valued Gram matrices (Panel b). We note that both $G_{opt}$ and $G_0$ are computed considering the input state permutation that results in the highest gain possible in each case. As we can see for both cases, the matrix $U_0$ results in a negative gain for most Gram matrix preparations. In particular, for the real matrices this is the case for around $91.6\%$ of the total preparations and for the complex matrices are around $87.4\%$. Notably, we can also see that in almost all cases, through the optical distillation matrix, it is possible to improve the overall gain when compared to that of the matrix $U_0$.
}


\begin{thebibliography}{54}%
\makeatletter
\providecommand \@ifxundefined [1]{%
 \@ifx{#1\undefined}
}%
\providecommand \@ifnum [1]{%
 \ifnum #1\expandafter \@firstoftwo
 \else \expandafter \@secondoftwo
 \fi
}%
\providecommand \@ifx [1]{%
 \ifx #1\expandafter \@firstoftwo
 \else \expandafter \@secondoftwo
 \fi
}%
\providecommand \natexlab [1]{#1}%
\providecommand \enquote  [1]{``#1''}%
\providecommand \bibnamefont  [1]{#1}%
\providecommand \bibfnamefont [1]{#1}%
\providecommand \citenamefont [1]{#1}%
\providecommand \href@noop [0]{\@secondoftwo}%
\providecommand \href [0]{\begingroup \@sanitize@url \@href}%
\providecommand \@href[1]{\@@startlink{#1}\@@href}%
\providecommand \@@href[1]{\endgroup#1\@@endlink}%
\providecommand \@sanitize@url [0]{\catcode `\\12\catcode `\$12\catcode `\&12\catcode `\#12\catcode `\^12\catcode `\_12\catcode `\%12\relax}%
\providecommand \@@startlink[1]{}%
\providecommand \@@endlink[0]{}%
\providecommand \url  [0]{\begingroup\@sanitize@url \@url }%
\providecommand \@url [1]{\endgroup\@href {#1}{\urlprefix }}%
\providecommand \urlprefix  [0]{URL }%
\providecommand \Eprint [0]{\href }%
\providecommand \doibase [0]{http://dx.doi.org/}%
\providecommand \selectlanguage [0]{\@gobble}%
\providecommand \bibinfo  [0]{\@secondoftwo}%
\providecommand \bibfield  [0]{\@secondoftwo}%
\providecommand \translation [1]{[#1]}%
\providecommand \BibitemOpen [0]{}%
\providecommand \bibitemStop [0]{}%
\providecommand \bibitemNoStop [0]{.\EOS\space}%
\providecommand \EOS [0]{\spacefactor3000\relax}%
\providecommand \BibitemShut  [1]{\csname bibitem#1\endcsname}%
\let\auto@bib@innerbib\@empty
\bibitem [{\citenamefont {Flamini}\ \emph {et~al.}(2018)\citenamefont {Flamini}, \citenamefont {Spagnolo},\ and\ \citenamefont {Sciarrino}}]{Flamini2018}%
  \BibitemOpen
  \bibfield  {author} {\bibinfo {author} {\bibfnamefont {F.}~\bibnamefont {Flamini}}, \bibinfo {author} {\bibfnamefont {N.}~\bibnamefont {Spagnolo}}, \ and\ \bibinfo {author} {\bibfnamefont {F.}~\bibnamefont {Sciarrino}},\ }\href {\doibase 10.1088/1361-6633/aad5b2} {\bibfield  {journal} {\bibinfo  {journal} {Reports on Progress in Physics}\ }\textbf {\bibinfo {volume} {82}},\ \bibinfo {pages} {016001} (\bibinfo {year} {2018})}\BibitemShut {NoStop}%
\bibitem [{\citenamefont {Knill}\ \emph {et~al.}(2001)\citenamefont {Knill}, \citenamefont {Laflamme},\ and\ \citenamefont {Milburn}}]{Knill2001}%
  \BibitemOpen
  \bibfield  {author} {\bibinfo {author} {\bibfnamefont {E.}~\bibnamefont {Knill}}, \bibinfo {author} {\bibfnamefont {R.}~\bibnamefont {Laflamme}}, \ and\ \bibinfo {author} {\bibfnamefont {G.~J.}\ \bibnamefont {Milburn}},\ }\href {\doibase 10.1038/35051009} {\bibfield  {journal} {\bibinfo  {journal} {Nature}\ }\textbf {\bibinfo {volume} {409}},\ \bibinfo {pages} {46–52} (\bibinfo {year} {2001})}\BibitemShut {NoStop}%
\bibitem [{\citenamefont {Kok}\ \emph {et~al.}(2007)\citenamefont {Kok}, \citenamefont {Munro}, \citenamefont {Nemoto}, \citenamefont {Ralph}, \citenamefont {Dowling},\ and\ \citenamefont {Milburn}}]{Kok_rev_2007}%
  \BibitemOpen
  \bibfield  {author} {\bibinfo {author} {\bibfnamefont {P.}~\bibnamefont {Kok}}, \bibinfo {author} {\bibfnamefont {W.~J.}\ \bibnamefont {Munro}}, \bibinfo {author} {\bibfnamefont {K.}~\bibnamefont {Nemoto}}, \bibinfo {author} {\bibfnamefont {T.~C.}\ \bibnamefont {Ralph}}, \bibinfo {author} {\bibfnamefont {J.~P.}\ \bibnamefont {Dowling}}, \ and\ \bibinfo {author} {\bibfnamefont {G.~J.}\ \bibnamefont {Milburn}},\ }\href {\doibase 10.1103/RevModPhys.79.135} {\bibfield  {journal} {\bibinfo  {journal} {Reviews of Modern Physics}\ }\textbf {\bibinfo {volume} {79}},\ \bibinfo {pages} {135} (\bibinfo {year} {2007})}\BibitemShut {NoStop}%
\bibitem [{\citenamefont {O'Brien}(2007)}]{Obrien_rev}%
  \BibitemOpen
  \bibfield  {author} {\bibinfo {author} {\bibfnamefont {J.~L.}\ \bibnamefont {O'Brien}},\ }\href {\doibase 10.1126/science.1142892} {\bibfield  {journal} {\bibinfo  {journal} {Science}\ }\textbf {\bibinfo {volume} {318}},\ \bibinfo {pages} {1567} (\bibinfo {year} {2007})}\BibitemShut {NoStop}%
\bibitem [{\citenamefont {Bouchard}\ \emph {et~al.}(2020)\citenamefont {Bouchard}, \citenamefont {Sit}, \citenamefont {Zhang}, \citenamefont {Fickler}, \citenamefont {Miatto}, \citenamefont {Yao}, \citenamefont {Sciarrino},\ and\ \citenamefont {Karimi}}]{Bouchard2020}%
  \BibitemOpen
  \bibfield  {author} {\bibinfo {author} {\bibfnamefont {F.}~\bibnamefont {Bouchard}}, \bibinfo {author} {\bibfnamefont {A.}~\bibnamefont {Sit}}, \bibinfo {author} {\bibfnamefont {Y.}~\bibnamefont {Zhang}}, \bibinfo {author} {\bibfnamefont {R.}~\bibnamefont {Fickler}}, \bibinfo {author} {\bibfnamefont {F.~M.}\ \bibnamefont {Miatto}}, \bibinfo {author} {\bibfnamefont {Y.}~\bibnamefont {Yao}}, \bibinfo {author} {\bibfnamefont {F.}~\bibnamefont {Sciarrino}}, \ and\ \bibinfo {author} {\bibfnamefont {E.}~\bibnamefont {Karimi}},\ }\href {\doibase 10.1088/1361-6633/abcd7a} {\bibfield  {journal} {\bibinfo  {journal} {Reports on Progress in Physics}\ }\textbf {\bibinfo {volume} {84}},\ \bibinfo {pages} {012402} (\bibinfo {year} {2020})}\BibitemShut {NoStop}%
\bibitem [{\citenamefont {Garcia-Escartin}\ and\ \citenamefont {Chamorro-Posada}(2013)}]{HOM_swap}%
  \BibitemOpen
  \bibfield  {author} {\bibinfo {author} {\bibfnamefont {J.~C.}\ \bibnamefont {Garcia-Escartin}}\ and\ \bibinfo {author} {\bibfnamefont {P.}~\bibnamefont {Chamorro-Posada}},\ }\href {\doibase 10.1103/PhysRevA.87.052330} {\bibfield  {journal} {\bibinfo  {journal} {Physical Review A}\ }\textbf {\bibinfo {volume} {87}},\ \bibinfo {pages} {052330} (\bibinfo {year} {2013})}\BibitemShut {NoStop}%
\bibitem [{\citenamefont {Briegel}\ \emph {et~al.}(2009)\citenamefont {Briegel}, \citenamefont {Browne}, \citenamefont {D\"{u}r}, \citenamefont {Raussendorf},\ and\ \citenamefont {Van~den Nest}}]{Briegel2009}%
  \BibitemOpen
  \bibfield  {author} {\bibinfo {author} {\bibfnamefont {H.~J.}\ \bibnamefont {Briegel}}, \bibinfo {author} {\bibfnamefont {D.~E.}\ \bibnamefont {Browne}}, \bibinfo {author} {\bibfnamefont {W.}~\bibnamefont {D\"{u}r}}, \bibinfo {author} {\bibfnamefont {R.}~\bibnamefont {Raussendorf}}, \ and\ \bibinfo {author} {\bibfnamefont {M.}~\bibnamefont {Van~den Nest}},\ }\href {\doibase 10.1038/nphys1157} {\bibfield  {journal} {\bibinfo  {journal} {Nature Physics}\ }\textbf {\bibinfo {volume} {5}},\ \bibinfo {pages} {19–26} (\bibinfo {year} {2009})}\BibitemShut {NoStop}%
\bibitem [{\citenamefont {Bartolucci}\ \emph {et~al.}(2023)\citenamefont {Bartolucci}, \citenamefont {Birchall}, \citenamefont {Bombín}, \citenamefont {Cable}, \citenamefont {Dawson}, \citenamefont {Gimeno-Segovia}, \citenamefont {Johnston}, \citenamefont {Kieling}, \citenamefont {Nickerson}, \citenamefont {Pant}, \citenamefont {Pastawski}, \citenamefont {Rudolph},\ and\ \citenamefont {Sparrow}}]{Bartolucci2023}%
  \BibitemOpen
  \bibfield  {author} {\bibinfo {author} {\bibfnamefont {S.}~\bibnamefont {Bartolucci}}, \bibinfo {author} {\bibfnamefont {P.}~\bibnamefont {Birchall}}, \bibinfo {author} {\bibfnamefont {H.}~\bibnamefont {Bombín}}, \bibinfo {author} {\bibfnamefont {H.}~\bibnamefont {Cable}}, \bibinfo {author} {\bibfnamefont {C.}~\bibnamefont {Dawson}}, \bibinfo {author} {\bibfnamefont {M.}~\bibnamefont {Gimeno-Segovia}}, \bibinfo {author} {\bibfnamefont {E.}~\bibnamefont {Johnston}}, \bibinfo {author} {\bibfnamefont {K.}~\bibnamefont {Kieling}}, \bibinfo {author} {\bibfnamefont {N.}~\bibnamefont {Nickerson}}, \bibinfo {author} {\bibfnamefont {M.}~\bibnamefont {Pant}}, \bibinfo {author} {\bibfnamefont {F.}~\bibnamefont {Pastawski}}, \bibinfo {author} {\bibfnamefont {T.}~\bibnamefont {Rudolph}}, \ and\ \bibinfo {author} {\bibfnamefont {C.}~\bibnamefont {Sparrow}},\ }\href {\doibase 10.1038/s41467-023-36493-1} {\bibfield  {journal} {\bibinfo  {journal} {Nature Communications}\ }\textbf {\bibinfo {volume} {14}},\ \bibinfo
  {pages} {912} (\bibinfo {year} {2023})}\BibitemShut {NoStop}%
\bibitem [{\citenamefont {Chen}\ \emph {et~al.}(2024)\citenamefont {Chen}, \citenamefont {Peng}, \citenamefont {Guo}, \citenamefont {Gu}, \citenamefont {Ding}, \citenamefont {Liu}, \citenamefont {Zhao}, \citenamefont {You}, \citenamefont {Qin}, \citenamefont {Wang}, \citenamefont {He}, \citenamefont {Renema}, \citenamefont {Huo}, \citenamefont {Wang}, \citenamefont {Lu},\ and\ \citenamefont {Pan}}]{Chen24}%
  \BibitemOpen
  \bibfield  {author} {\bibinfo {author} {\bibfnamefont {S.}~\bibnamefont {Chen}}, \bibinfo {author} {\bibfnamefont {L.-C.}\ \bibnamefont {Peng}}, \bibinfo {author} {\bibfnamefont {Y.-P.}\ \bibnamefont {Guo}}, \bibinfo {author} {\bibfnamefont {X.-M.}\ \bibnamefont {Gu}}, \bibinfo {author} {\bibfnamefont {X.}~\bibnamefont {Ding}}, \bibinfo {author} {\bibfnamefont {R.-Z.}\ \bibnamefont {Liu}}, \bibinfo {author} {\bibfnamefont {J.-Y.}\ \bibnamefont {Zhao}}, \bibinfo {author} {\bibfnamefont {X.}~\bibnamefont {You}}, \bibinfo {author} {\bibfnamefont {J.}~\bibnamefont {Qin}}, \bibinfo {author} {\bibfnamefont {Y.-F.}\ \bibnamefont {Wang}}, \bibinfo {author} {\bibfnamefont {Y.-M.}\ \bibnamefont {He}}, \bibinfo {author} {\bibfnamefont {J.~J.}\ \bibnamefont {Renema}}, \bibinfo {author} {\bibfnamefont {Y.-H.}\ \bibnamefont {Huo}}, \bibinfo {author} {\bibfnamefont {H.}~\bibnamefont {Wang}}, \bibinfo {author} {\bibfnamefont {C.-Y.}\ \bibnamefont {Lu}}, \ and\ \bibinfo {author} {\bibfnamefont {J.-W.}\ \bibnamefont {Pan}},\
  }\href {\doibase 10.1103/PhysRevLett.132.130603} {\bibfield  {journal} {\bibinfo  {journal} {Physical Review Letters}\ }\textbf {\bibinfo {volume} {132}},\ \bibinfo {pages} {130603} (\bibinfo {year} {2024})}\BibitemShut {NoStop}%
\bibitem [{\citenamefont {Cao}\ \emph {et~al.}(2024)\citenamefont {Cao}, \citenamefont {Hansen}, \citenamefont {Giorgino}, \citenamefont {Carosini}, \citenamefont {Zah\'alka}, \citenamefont {Zilk}, \citenamefont {Loredo},\ and\ \citenamefont {Walther}}]{Cao24}%
  \BibitemOpen
  \bibfield  {author} {\bibinfo {author} {\bibfnamefont {H.}~\bibnamefont {Cao}}, \bibinfo {author} {\bibfnamefont {L.~M.}\ \bibnamefont {Hansen}}, \bibinfo {author} {\bibfnamefont {F.}~\bibnamefont {Giorgino}}, \bibinfo {author} {\bibfnamefont {L.}~\bibnamefont {Carosini}}, \bibinfo {author} {\bibfnamefont {P.}~\bibnamefont {Zah\'alka}}, \bibinfo {author} {\bibfnamefont {F.}~\bibnamefont {Zilk}}, \bibinfo {author} {\bibfnamefont {J.~C.}\ \bibnamefont {Loredo}}, \ and\ \bibinfo {author} {\bibfnamefont {P.}~\bibnamefont {Walther}},\ }\href {\doibase 10.1103/PhysRevLett.132.130604} {\bibfield  {journal} {\bibinfo  {journal} {Physical Review Letters}\ }\textbf {\bibinfo {volume} {132}},\ \bibinfo {pages} {130604} (\bibinfo {year} {2024})}\BibitemShut {NoStop}%
\bibitem [{\citenamefont {Pont}\ \emph {et~al.}(2024)\citenamefont {Pont}, \citenamefont {Corrielli}, \citenamefont {Fyrillas}, \citenamefont {Agresti}, \citenamefont {Carvacho}, \citenamefont {Maring}, \citenamefont {Emeriau}, \citenamefont {Ceccarelli}, \citenamefont {Albiero}, \citenamefont {Dias~Ferreira}, \citenamefont {Somaschi}, \citenamefont {Senellart}, \citenamefont {Sagnes}, \citenamefont {Morassi}, \citenamefont {Lemaître}, \citenamefont {Senellart}, \citenamefont {Sciarrino}, \citenamefont {Liscidini}, \citenamefont {Belabas},\ and\ \citenamefont {Osellame}}]{Pont2024}%
  \BibitemOpen
  \bibfield  {author} {\bibinfo {author} {\bibfnamefont {M.}~\bibnamefont {Pont}}, \bibinfo {author} {\bibfnamefont {G.}~\bibnamefont {Corrielli}}, \bibinfo {author} {\bibfnamefont {A.}~\bibnamefont {Fyrillas}}, \bibinfo {author} {\bibfnamefont {I.}~\bibnamefont {Agresti}}, \bibinfo {author} {\bibfnamefont {G.}~\bibnamefont {Carvacho}}, \bibinfo {author} {\bibfnamefont {N.}~\bibnamefont {Maring}}, \bibinfo {author} {\bibfnamefont {P.-E.}\ \bibnamefont {Emeriau}}, \bibinfo {author} {\bibfnamefont {F.}~\bibnamefont {Ceccarelli}}, \bibinfo {author} {\bibfnamefont {R.}~\bibnamefont {Albiero}}, \bibinfo {author} {\bibfnamefont {P.~H.}\ \bibnamefont {Dias~Ferreira}}, \bibinfo {author} {\bibfnamefont {N.}~\bibnamefont {Somaschi}}, \bibinfo {author} {\bibfnamefont {J.}~\bibnamefont {Senellart}}, \bibinfo {author} {\bibfnamefont {I.}~\bibnamefont {Sagnes}}, \bibinfo {author} {\bibfnamefont {M.}~\bibnamefont {Morassi}}, \bibinfo {author} {\bibfnamefont {A.}~\bibnamefont {Lemaître}}, \bibinfo {author} {\bibfnamefont
  {P.}~\bibnamefont {Senellart}}, \bibinfo {author} {\bibfnamefont {F.}~\bibnamefont {Sciarrino}}, \bibinfo {author} {\bibfnamefont {M.}~\bibnamefont {Liscidini}}, \bibinfo {author} {\bibfnamefont {N.}~\bibnamefont {Belabas}}, \ and\ \bibinfo {author} {\bibfnamefont {R.}~\bibnamefont {Osellame}},\ }\href {\doibase 10.1038/s41534-024-00830-z} {\bibfield  {journal} {\bibinfo  {journal} {npj Quantum Information}\ }\textbf {\bibinfo {volume} {10}},\ \bibinfo {pages} {50} (\bibinfo {year} {2024})}\BibitemShut {NoStop}%
\bibitem [{\citenamefont {Gisin}\ and\ \citenamefont {Thew}(2007)}]{Gisin2007}%
  \BibitemOpen
  \bibfield  {author} {\bibinfo {author} {\bibfnamefont {N.}~\bibnamefont {Gisin}}\ and\ \bibinfo {author} {\bibfnamefont {R.}~\bibnamefont {Thew}},\ }\href {\doibase 10.1038/nphoton.2007.22} {\bibfield  {journal} {\bibinfo  {journal} {Nature Photonics}\ }\textbf {\bibinfo {volume} {1}},\ \bibinfo {pages} {165–171} (\bibinfo {year} {2007})}\BibitemShut {NoStop}%
\bibitem [{\citenamefont {Pirandola}\ \emph {et~al.}(2015)\citenamefont {Pirandola}, \citenamefont {Eisert}, \citenamefont {Weedbrook}, \citenamefont {Furusawa},\ and\ \citenamefont {Braunstein}}]{Pirandola2015}%
  \BibitemOpen
  \bibfield  {author} {\bibinfo {author} {\bibfnamefont {S.}~\bibnamefont {Pirandola}}, \bibinfo {author} {\bibfnamefont {J.}~\bibnamefont {Eisert}}, \bibinfo {author} {\bibfnamefont {C.}~\bibnamefont {Weedbrook}}, \bibinfo {author} {\bibfnamefont {A.}~\bibnamefont {Furusawa}}, \ and\ \bibinfo {author} {\bibfnamefont {S.~L.}\ \bibnamefont {Braunstein}},\ }\href {\doibase 10.1038/nphoton.2015.154} {\bibfield  {journal} {\bibinfo  {journal} {Nature Photonics}\ }\textbf {\bibinfo {volume} {9}},\ \bibinfo {pages} {641–652} (\bibinfo {year} {2015})}\BibitemShut {NoStop}%
\bibitem [{\citenamefont {Hu}\ \emph {et~al.}(2023)\citenamefont {Hu}, \citenamefont {Guo}, \citenamefont {Liu}, \citenamefont {Li},\ and\ \citenamefont {Guo}}]{Hu2023}%
  \BibitemOpen
  \bibfield  {author} {\bibinfo {author} {\bibfnamefont {X.-M.}\ \bibnamefont {Hu}}, \bibinfo {author} {\bibfnamefont {Y.}~\bibnamefont {Guo}}, \bibinfo {author} {\bibfnamefont {B.-H.}\ \bibnamefont {Liu}}, \bibinfo {author} {\bibfnamefont {C.-F.}\ \bibnamefont {Li}}, \ and\ \bibinfo {author} {\bibfnamefont {G.-C.}\ \bibnamefont {Guo}},\ }\href {\doibase 10.1038/s42254-023-00588-x} {\bibfield  {journal} {\bibinfo  {journal} {Nature Reviews Physics}\ }\textbf {\bibinfo {volume} {5}},\ \bibinfo {pages} {339–353} (\bibinfo {year} {2023})}\BibitemShut {NoStop}%
\bibitem [{\citenamefont {Carvacho}\ \emph {et~al.}(2017)\citenamefont {Carvacho}, \citenamefont {Andreoli}, \citenamefont {Santodonato}, \citenamefont {Bentivegna}, \citenamefont {Chaves},\ and\ \citenamefont {Sciarrino}}]{Carvacho2017}%
  \BibitemOpen
  \bibfield  {author} {\bibinfo {author} {\bibfnamefont {G.}~\bibnamefont {Carvacho}}, \bibinfo {author} {\bibfnamefont {F.}~\bibnamefont {Andreoli}}, \bibinfo {author} {\bibfnamefont {L.}~\bibnamefont {Santodonato}}, \bibinfo {author} {\bibfnamefont {M.}~\bibnamefont {Bentivegna}}, \bibinfo {author} {\bibfnamefont {R.}~\bibnamefont {Chaves}}, \ and\ \bibinfo {author} {\bibfnamefont {F.}~\bibnamefont {Sciarrino}},\ }\href {\doibase 10.1038/ncomms14775} {\bibfield  {journal} {\bibinfo  {journal} {Nature Communications}\ }\textbf {\bibinfo {volume} {8}},\ \bibinfo {pages} {14775} (\bibinfo {year} {2017})}\BibitemShut {NoStop}%
\bibitem [{\citenamefont {Aaronson}\ and\ \citenamefont {Arkhipov}(2011)}]{AA_2010}%
  \BibitemOpen
  \bibfield  {author} {\bibinfo {author} {\bibfnamefont {S.}~\bibnamefont {Aaronson}}\ and\ \bibinfo {author} {\bibfnamefont {A.}~\bibnamefont {Arkhipov}},\ }in\ \href {\doibase 10.1145/1993636.1993682} {\emph {\bibinfo {booktitle} {Proceedings of the Forty-Third Annual ACM Symposium on Theory of Computing}}},\ \bibinfo {series and number} {STOC '11}\ (\bibinfo  {publisher} {Association for Computing Machinery},\ \bibinfo {address} {New York, NY, USA},\ \bibinfo {year} {2011})\ p.\ \bibinfo {pages} {333–342}\BibitemShut {NoStop}%
\bibitem [{\citenamefont {Brod}\ \emph {et~al.}(2019{\natexlab{a}})\citenamefont {Brod}, \citenamefont {Galv{\~a}o}, \citenamefont {Crespi}, \citenamefont {Osellame}, \citenamefont {Spagnolo},\ and\ \citenamefont {Sciarrino}}]{Brod_rev}%
  \BibitemOpen
  \bibfield  {author} {\bibinfo {author} {\bibfnamefont {D.~J.}\ \bibnamefont {Brod}}, \bibinfo {author} {\bibfnamefont {E.~F.}\ \bibnamefont {Galv{\~a}o}}, \bibinfo {author} {\bibfnamefont {A.}~\bibnamefont {Crespi}}, \bibinfo {author} {\bibfnamefont {R.}~\bibnamefont {Osellame}}, \bibinfo {author} {\bibfnamefont {N.}~\bibnamefont {Spagnolo}}, \ and\ \bibinfo {author} {\bibfnamefont {F.}~\bibnamefont {Sciarrino}},\ }\href {\doibase 10.1117/1.AP.1.3.034001} {\bibfield  {journal} {\bibinfo  {journal} {Advanced Photonics}\ }\textbf {\bibinfo {volume} {1}},\ \bibinfo {pages} {034001} (\bibinfo {year} {2019}{\natexlab{a}})}\BibitemShut {NoStop}%
\bibitem [{\citenamefont {Monbroussou}\ \emph {et~al.}(2025)\citenamefont {Monbroussou}, \citenamefont {Polacchi}, \citenamefont {Yacoub}, \citenamefont {Caruccio}, \citenamefont {Rodari}, \citenamefont {Hoch}, \citenamefont {Carvacho}, \citenamefont {Spagnolo}, \citenamefont {Giordani}, \citenamefont {Bossi}, \citenamefont {Rajan}, \citenamefont {Giano}, \citenamefont {Albiero}, \citenamefont {Ceccarelli}, \citenamefont {Osellame}, \citenamefont {Kashefi},\ and\ \citenamefont {Sciarrino}}]{polacchi2025}%
  \BibitemOpen
  \bibfield  {author} {\bibinfo {author} {\bibfnamefont {L.}~\bibnamefont {Monbroussou}}, \bibinfo {author} {\bibfnamefont {B.}~\bibnamefont {Polacchi}}, \bibinfo {author} {\bibfnamefont {V.}~\bibnamefont {Yacoub}}, \bibinfo {author} {\bibfnamefont {E.}~\bibnamefont {Caruccio}}, \bibinfo {author} {\bibfnamefont {G.}~\bibnamefont {Rodari}}, \bibinfo {author} {\bibfnamefont {F.}~\bibnamefont {Hoch}}, \bibinfo {author} {\bibfnamefont {G.}~\bibnamefont {Carvacho}}, \bibinfo {author} {\bibfnamefont {N.}~\bibnamefont {Spagnolo}}, \bibinfo {author} {\bibfnamefont {T.}~\bibnamefont {Giordani}}, \bibinfo {author} {\bibfnamefont {M.}~\bibnamefont {Bossi}}, \bibinfo {author} {\bibfnamefont {A.}~\bibnamefont {Rajan}}, \bibinfo {author} {\bibfnamefont {N.~D.}\ \bibnamefont {Giano}}, \bibinfo {author} {\bibfnamefont {R.}~\bibnamefont {Albiero}}, \bibinfo {author} {\bibfnamefont {F.}~\bibnamefont {Ceccarelli}}, \bibinfo {author} {\bibfnamefont {R.}~\bibnamefont {Osellame}}, \bibinfo {author} {\bibfnamefont {E.}~\bibnamefont
  {Kashefi}}, \ and\ \bibinfo {author} {\bibfnamefont {F.}~\bibnamefont {Sciarrino}},\ }\href {https://arxiv.org/abs/2504.20989} {\enquote {\bibinfo {title} {Photonic quantum convolutional neural networks with adaptive state injection},}\ } (\bibinfo {year} {2025}),\ \Eprint {http://arxiv.org/abs/2504.20989} {arXiv:2504.20989 [quant-ph]} \BibitemShut {NoStop}%
\bibitem [{\citenamefont {Hoch}\ \emph {et~al.}(2025)\citenamefont {Hoch}, \citenamefont {Caruccio}, \citenamefont {Rodari}, \citenamefont {Francalanci}, \citenamefont {Suprano}, \citenamefont {Giordani}, \citenamefont {Carvacho}, \citenamefont {Spagnolo}, \citenamefont {Koudia}, \citenamefont {Proietti}, \citenamefont {Liorni}, \citenamefont {Cerocchi}, \citenamefont {Albiero}, \citenamefont {Di~Giano}, \citenamefont {Gardina}, \citenamefont {Ceccarelli}, \citenamefont {Corrielli}, \citenamefont {Chabaud}, \citenamefont {Osellame}, \citenamefont {Dispenza},\ and\ \citenamefont {Sciarrino}}]{Hoch2025}%
  \BibitemOpen
  \bibfield  {author} {\bibinfo {author} {\bibfnamefont {F.}~\bibnamefont {Hoch}}, \bibinfo {author} {\bibfnamefont {E.}~\bibnamefont {Caruccio}}, \bibinfo {author} {\bibfnamefont {G.}~\bibnamefont {Rodari}}, \bibinfo {author} {\bibfnamefont {T.}~\bibnamefont {Francalanci}}, \bibinfo {author} {\bibfnamefont {A.}~\bibnamefont {Suprano}}, \bibinfo {author} {\bibfnamefont {T.}~\bibnamefont {Giordani}}, \bibinfo {author} {\bibfnamefont {G.}~\bibnamefont {Carvacho}}, \bibinfo {author} {\bibfnamefont {N.}~\bibnamefont {Spagnolo}}, \bibinfo {author} {\bibfnamefont {S.}~\bibnamefont {Koudia}}, \bibinfo {author} {\bibfnamefont {M.}~\bibnamefont {Proietti}}, \bibinfo {author} {\bibfnamefont {C.}~\bibnamefont {Liorni}}, \bibinfo {author} {\bibfnamefont {F.}~\bibnamefont {Cerocchi}}, \bibinfo {author} {\bibfnamefont {R.}~\bibnamefont {Albiero}}, \bibinfo {author} {\bibfnamefont {N.}~\bibnamefont {Di~Giano}}, \bibinfo {author} {\bibfnamefont {M.}~\bibnamefont {Gardina}}, \bibinfo {author} {\bibfnamefont {F.}~\bibnamefont
  {Ceccarelli}}, \bibinfo {author} {\bibfnamefont {G.}~\bibnamefont {Corrielli}}, \bibinfo {author} {\bibfnamefont {U.}~\bibnamefont {Chabaud}}, \bibinfo {author} {\bibfnamefont {R.}~\bibnamefont {Osellame}}, \bibinfo {author} {\bibfnamefont {M.}~\bibnamefont {Dispenza}}, \ and\ \bibinfo {author} {\bibfnamefont {F.}~\bibnamefont {Sciarrino}},\ }\href {\doibase 10.1038/s41467-025-55877-z} {\bibfield  {journal} {\bibinfo  {journal} {Nature Communications}\ }\textbf {\bibinfo {volume} {16}},\ \bibinfo {pages} {902} (\bibinfo {year} {2025})}\BibitemShut {NoStop}%
\bibitem [{\citenamefont {Spagnolo}\ \emph {et~al.}(2013)\citenamefont {Spagnolo}, \citenamefont {Vitelli}, \citenamefont {Aparo}, \citenamefont {Mataloni}, \citenamefont {Sciarrino}, \citenamefont {Crespi}, \citenamefont {Ramponi},\ and\ \citenamefont {Osellame}}]{Spagnolo2013}%
  \BibitemOpen
  \bibfield  {author} {\bibinfo {author} {\bibfnamefont {N.}~\bibnamefont {Spagnolo}}, \bibinfo {author} {\bibfnamefont {C.}~\bibnamefont {Vitelli}}, \bibinfo {author} {\bibfnamefont {L.}~\bibnamefont {Aparo}}, \bibinfo {author} {\bibfnamefont {P.}~\bibnamefont {Mataloni}}, \bibinfo {author} {\bibfnamefont {F.}~\bibnamefont {Sciarrino}}, \bibinfo {author} {\bibfnamefont {A.}~\bibnamefont {Crespi}}, \bibinfo {author} {\bibfnamefont {R.}~\bibnamefont {Ramponi}}, \ and\ \bibinfo {author} {\bibfnamefont {R.}~\bibnamefont {Osellame}},\ }\href {\doibase 10.1038/ncomms2616} {\bibfield  {journal} {\bibinfo  {journal} {Nature Communications}\ }\textbf {\bibinfo {volume} {4}},\ \bibinfo {pages} {1606} (\bibinfo {year} {2013})}\BibitemShut {NoStop}%
\bibitem [{\citenamefont {Menssen}\ \emph {et~al.}(2017)\citenamefont {Menssen}, \citenamefont {Jones}, \citenamefont {Metcalf}, \citenamefont {Tichy}, \citenamefont {Barz}, \citenamefont {Kolthammer},\ and\ \citenamefont {Walmsley}}]{Menssen_17}%
  \BibitemOpen
  \bibfield  {author} {\bibinfo {author} {\bibfnamefont {A.~J.}\ \bibnamefont {Menssen}}, \bibinfo {author} {\bibfnamefont {A.~E.}\ \bibnamefont {Jones}}, \bibinfo {author} {\bibfnamefont {B.~J.}\ \bibnamefont {Metcalf}}, \bibinfo {author} {\bibfnamefont {M.~C.}\ \bibnamefont {Tichy}}, \bibinfo {author} {\bibfnamefont {S.}~\bibnamefont {Barz}}, \bibinfo {author} {\bibfnamefont {W.~S.}\ \bibnamefont {Kolthammer}}, \ and\ \bibinfo {author} {\bibfnamefont {I.~A.}\ \bibnamefont {Walmsley}},\ }\href {\doibase 10.1103/PhysRevLett.118.153603} {\bibfield  {journal} {\bibinfo  {journal} {Physical Review Letters}\ }\textbf {\bibinfo {volume} {118}},\ \bibinfo {pages} {153603} (\bibinfo {year} {2017})}\BibitemShut {NoStop}%
\bibitem [{\citenamefont {Agne}\ \emph {et~al.}(2017)\citenamefont {Agne}, \citenamefont {Kauten}, \citenamefont {Jin}, \citenamefont {Meyer-Scott}, \citenamefont {Salvail}, \citenamefont {Hamel}, \citenamefont {Resch}, \citenamefont {Weihs},\ and\ \citenamefont {Jennewein}}]{Agne_17}%
  \BibitemOpen
  \bibfield  {author} {\bibinfo {author} {\bibfnamefont {S.}~\bibnamefont {Agne}}, \bibinfo {author} {\bibfnamefont {T.}~\bibnamefont {Kauten}}, \bibinfo {author} {\bibfnamefont {J.}~\bibnamefont {Jin}}, \bibinfo {author} {\bibfnamefont {E.}~\bibnamefont {Meyer-Scott}}, \bibinfo {author} {\bibfnamefont {J.~Z.}\ \bibnamefont {Salvail}}, \bibinfo {author} {\bibfnamefont {D.~R.}\ \bibnamefont {Hamel}}, \bibinfo {author} {\bibfnamefont {K.~J.}\ \bibnamefont {Resch}}, \bibinfo {author} {\bibfnamefont {G.}~\bibnamefont {Weihs}}, \ and\ \bibinfo {author} {\bibfnamefont {T.}~\bibnamefont {Jennewein}},\ }\href {\doibase 10.1103/PhysRevLett.118.153602} {\bibfield  {journal} {\bibinfo  {journal} {Physical Review Letters}\ }\textbf {\bibinfo {volume} {118}},\ \bibinfo {pages} {153602} (\bibinfo {year} {2017})}\BibitemShut {NoStop}%
\bibitem [{\citenamefont {Brod}\ \emph {et~al.}(2019{\natexlab{b}})\citenamefont {Brod}, \citenamefont {Galv\~ao}, \citenamefont {Viggianiello}, \citenamefont {Flamini}, \citenamefont {Spagnolo},\ and\ \citenamefont {Sciarrino}}]{Brod_19}%
  \BibitemOpen
  \bibfield  {author} {\bibinfo {author} {\bibfnamefont {D.~J.}\ \bibnamefont {Brod}}, \bibinfo {author} {\bibfnamefont {E.~F.}\ \bibnamefont {Galv\~ao}}, \bibinfo {author} {\bibfnamefont {N.}~\bibnamefont {Viggianiello}}, \bibinfo {author} {\bibfnamefont {F.}~\bibnamefont {Flamini}}, \bibinfo {author} {\bibfnamefont {N.}~\bibnamefont {Spagnolo}}, \ and\ \bibinfo {author} {\bibfnamefont {F.}~\bibnamefont {Sciarrino}},\ }\href {\doibase 10.1103/PhysRevLett.122.063602} {\bibfield  {journal} {\bibinfo  {journal} {Physical Review Letters}\ }\textbf {\bibinfo {volume} {122}},\ \bibinfo {pages} {063602} (\bibinfo {year} {2019}{\natexlab{b}})}\BibitemShut {NoStop}%
\bibitem [{\citenamefont {Giordani}\ \emph {et~al.}(2020)\citenamefont {Giordani}, \citenamefont {Brod}, \citenamefont {Esposito}, \citenamefont {Viggianiello}, \citenamefont {Romano}, \citenamefont {Flamini}, \citenamefont {Carvacho}, \citenamefont {Spagnolo}, \citenamefont {Galvão},\ and\ \citenamefont {Sciarrino}}]{Giordani2020}%
  \BibitemOpen
  \bibfield  {author} {\bibinfo {author} {\bibfnamefont {T.}~\bibnamefont {Giordani}}, \bibinfo {author} {\bibfnamefont {D.~J.}\ \bibnamefont {Brod}}, \bibinfo {author} {\bibfnamefont {C.}~\bibnamefont {Esposito}}, \bibinfo {author} {\bibfnamefont {N.}~\bibnamefont {Viggianiello}}, \bibinfo {author} {\bibfnamefont {M.}~\bibnamefont {Romano}}, \bibinfo {author} {\bibfnamefont {F.}~\bibnamefont {Flamini}}, \bibinfo {author} {\bibfnamefont {G.}~\bibnamefont {Carvacho}}, \bibinfo {author} {\bibfnamefont {N.}~\bibnamefont {Spagnolo}}, \bibinfo {author} {\bibfnamefont {E.~F.}\ \bibnamefont {Galvão}}, \ and\ \bibinfo {author} {\bibfnamefont {F.}~\bibnamefont {Sciarrino}},\ }\href {\doibase 10.1088/1367-2630/ab7a30} {\bibfield  {journal} {\bibinfo  {journal} {New Journal of Physics}\ }\textbf {\bibinfo {volume} {22}},\ \bibinfo {pages} {043001} (\bibinfo {year} {2020})}\BibitemShut {NoStop}%
\bibitem [{\citenamefont {Pont}\ \emph {et~al.}(2022)\citenamefont {Pont}, \citenamefont {Albiero}, \citenamefont {Thomas}, \citenamefont {Spagnolo}, \citenamefont {Ceccarelli}, \citenamefont {Corrielli}, \citenamefont {Brieussel}, \citenamefont {Somaschi}, \citenamefont {Huet}, \citenamefont {Harouri}, \citenamefont {Lema\^{\i}tre}, \citenamefont {Sagnes}, \citenamefont {Belabas}, \citenamefont {Sciarrino}, \citenamefont {Osellame}, \citenamefont {Senellart},\ and\ \citenamefont {Crespi}}]{Pont_22}%
  \BibitemOpen
  \bibfield  {author} {\bibinfo {author} {\bibfnamefont {M.}~\bibnamefont {Pont}}, \bibinfo {author} {\bibfnamefont {R.}~\bibnamefont {Albiero}}, \bibinfo {author} {\bibfnamefont {S.~E.}\ \bibnamefont {Thomas}}, \bibinfo {author} {\bibfnamefont {N.}~\bibnamefont {Spagnolo}}, \bibinfo {author} {\bibfnamefont {F.}~\bibnamefont {Ceccarelli}}, \bibinfo {author} {\bibfnamefont {G.}~\bibnamefont {Corrielli}}, \bibinfo {author} {\bibfnamefont {A.}~\bibnamefont {Brieussel}}, \bibinfo {author} {\bibfnamefont {N.}~\bibnamefont {Somaschi}}, \bibinfo {author} {\bibfnamefont {H.}~\bibnamefont {Huet}}, \bibinfo {author} {\bibfnamefont {A.}~\bibnamefont {Harouri}}, \bibinfo {author} {\bibfnamefont {A.}~\bibnamefont {Lema\^{\i}tre}}, \bibinfo {author} {\bibfnamefont {I.}~\bibnamefont {Sagnes}}, \bibinfo {author} {\bibfnamefont {N.}~\bibnamefont {Belabas}}, \bibinfo {author} {\bibfnamefont {F.}~\bibnamefont {Sciarrino}}, \bibinfo {author} {\bibfnamefont {R.}~\bibnamefont {Osellame}}, \bibinfo {author} {\bibfnamefont
  {P.}~\bibnamefont {Senellart}}, \ and\ \bibinfo {author} {\bibfnamefont {A.}~\bibnamefont {Crespi}},\ }\href {\doibase 10.1103/PhysRevX.12.031033} {\bibfield  {journal} {\bibinfo  {journal} {Physical Review X}\ }\textbf {\bibinfo {volume} {12}},\ \bibinfo {pages} {031033} (\bibinfo {year} {2022})}\BibitemShut {NoStop}%
\bibitem [{\citenamefont {Seron}\ \emph {et~al.}(2023)\citenamefont {Seron}, \citenamefont {Novo},\ and\ \citenamefont {Cerf}}]{Seron2023}%
  \BibitemOpen
  \bibfield  {author} {\bibinfo {author} {\bibfnamefont {B.}~\bibnamefont {Seron}}, \bibinfo {author} {\bibfnamefont {L.}~\bibnamefont {Novo}}, \ and\ \bibinfo {author} {\bibfnamefont {N.~J.}\ \bibnamefont {Cerf}},\ }\href {\doibase 10.1038/s41566-023-01213-0} {\bibfield  {journal} {\bibinfo  {journal} {Nature Photonics}\ }\textbf {\bibinfo {volume} {17}},\ \bibinfo {pages} {702–709} (\bibinfo {year} {2023})}\BibitemShut {NoStop}%
\bibitem [{\citenamefont {Rodari}\ \emph {et~al.}(2025)\citenamefont {Rodari}, \citenamefont {Francalanci}, \citenamefont {Caruccio}, \citenamefont {Hoch}, \citenamefont {Carvacho}, \citenamefont {Giordani}, \citenamefont {Spagnolo}, \citenamefont {Albiero}, \citenamefont {Di~Giano}, \citenamefont {Ceccarelli}, \citenamefont {Corrielli}, \citenamefont {Crespi}, \citenamefont {Osellame}, \citenamefont {Chabaud},\ and\ \citenamefont {Sciarrino}}]{Rodari2025LAI}%
  \BibitemOpen
  \bibfield  {author} {\bibinfo {author} {\bibfnamefont {G.}~\bibnamefont {Rodari}}, \bibinfo {author} {\bibfnamefont {T.}~\bibnamefont {Francalanci}}, \bibinfo {author} {\bibfnamefont {E.}~\bibnamefont {Caruccio}}, \bibinfo {author} {\bibfnamefont {F.}~\bibnamefont {Hoch}}, \bibinfo {author} {\bibfnamefont {G.}~\bibnamefont {Carvacho}}, \bibinfo {author} {\bibfnamefont {T.}~\bibnamefont {Giordani}}, \bibinfo {author} {\bibfnamefont {N.}~\bibnamefont {Spagnolo}}, \bibinfo {author} {\bibfnamefont {R.}~\bibnamefont {Albiero}}, \bibinfo {author} {\bibfnamefont {N.}~\bibnamefont {Di~Giano}}, \bibinfo {author} {\bibfnamefont {F.}~\bibnamefont {Ceccarelli}}, \bibinfo {author} {\bibfnamefont {G.}~\bibnamefont {Corrielli}}, \bibinfo {author} {\bibfnamefont {A.}~\bibnamefont {Crespi}}, \bibinfo {author} {\bibfnamefont {R.}~\bibnamefont {Osellame}}, \bibinfo {author} {\bibfnamefont {U.}~\bibnamefont {Chabaud}}, \ and\ \bibinfo {author} {\bibfnamefont {F.}~\bibnamefont {Sciarrino}},\ }\href
  {https://arxiv.org/abs/2505.03001} {\enquote {\bibinfo {title} {Observation of lie algebraic invariants in quantum linear optics},}\ } (\bibinfo {year} {2025}),\ \Eprint {http://arxiv.org/abs/2505.03001} {arXiv:2505.03001 [quant-ph]} \BibitemShut {NoStop}%
\bibitem [{\citenamefont {Sund}\ \emph {et~al.}(2024)\citenamefont {Sund}, \citenamefont {Uppu}, \citenamefont {Paesani},\ and\ \citenamefont {Lodahl}}]{Sund_24}%
  \BibitemOpen
  \bibfield  {author} {\bibinfo {author} {\bibfnamefont {P.~I.}\ \bibnamefont {Sund}}, \bibinfo {author} {\bibfnamefont {R.}~\bibnamefont {Uppu}}, \bibinfo {author} {\bibfnamefont {S.}~\bibnamefont {Paesani}}, \ and\ \bibinfo {author} {\bibfnamefont {P.}~\bibnamefont {Lodahl}},\ }\href {\doibase 10.1103/PhysRevA.109.042613} {\bibfield  {journal} {\bibinfo  {journal} {Physical Review A}\ }\textbf {\bibinfo {volume} {109}},\ \bibinfo {pages} {042613} (\bibinfo {year} {2024})}\BibitemShut {NoStop}%
\bibitem [{\citenamefont {Senellart}\ \emph {et~al.}(2017)\citenamefont {Senellart}, \citenamefont {Solomon},\ and\ \citenamefont {White}}]{Senellart2017}%
  \BibitemOpen
  \bibfield  {author} {\bibinfo {author} {\bibfnamefont {P.}~\bibnamefont {Senellart}}, \bibinfo {author} {\bibfnamefont {G.}~\bibnamefont {Solomon}}, \ and\ \bibinfo {author} {\bibfnamefont {A.}~\bibnamefont {White}},\ }\href {\doibase 10.1038/nnano.2017.218} {\bibfield  {journal} {\bibinfo  {journal} {Nature Nanotechnology}\ }\textbf {\bibinfo {volume} {12}},\ \bibinfo {pages} {1026–1039} (\bibinfo {year} {2017})}\BibitemShut {NoStop}%
\bibitem [{\citenamefont {Caspani}\ \emph {et~al.}(2017)\citenamefont {Caspani}, \citenamefont {Xiong}, \citenamefont {Eggleton}, \citenamefont {Bajoni}, \citenamefont {Liscidini}, \citenamefont {Galli}, \citenamefont {Morandotti},\ and\ \citenamefont {Moss}}]{Caspani2017}%
  \BibitemOpen
  \bibfield  {author} {\bibinfo {author} {\bibfnamefont {L.}~\bibnamefont {Caspani}}, \bibinfo {author} {\bibfnamefont {C.}~\bibnamefont {Xiong}}, \bibinfo {author} {\bibfnamefont {B.~J.}\ \bibnamefont {Eggleton}}, \bibinfo {author} {\bibfnamefont {D.}~\bibnamefont {Bajoni}}, \bibinfo {author} {\bibfnamefont {M.}~\bibnamefont {Liscidini}}, \bibinfo {author} {\bibfnamefont {M.}~\bibnamefont {Galli}}, \bibinfo {author} {\bibfnamefont {R.}~\bibnamefont {Morandotti}}, \ and\ \bibinfo {author} {\bibfnamefont {D.~J.}\ \bibnamefont {Moss}},\ }\href {\doibase 10.1038/lsa.2017.100} {\bibfield  {journal} {\bibinfo  {journal} {Light: Science and Applications}\ }\textbf {\bibinfo {volume} {6}},\ \bibinfo {pages} {e17100} (\bibinfo {year} {2017})}\BibitemShut {NoStop}%
\bibitem [{\citenamefont {Somaschi}\ \emph {et~al.}(2016)\citenamefont {Somaschi}, \citenamefont {Giesz}, \citenamefont {De~Santis}, \citenamefont {Loredo}, \citenamefont {Almeida}, \citenamefont {Hornecker}, \citenamefont {Portalupi}, \citenamefont {Grange}, \citenamefont {Antón}, \citenamefont {Demory}, \citenamefont {Gómez}, \citenamefont {Sagnes}, \citenamefont {Lanzillotti-Kimura}, \citenamefont {Lemaítre}, \citenamefont {Auffeves}, \citenamefont {White}, \citenamefont {Lanco},\ and\ \citenamefont {Senellart}}]{Somaschi2016}%
  \BibitemOpen
  \bibfield  {author} {\bibinfo {author} {\bibfnamefont {N.}~\bibnamefont {Somaschi}}, \bibinfo {author} {\bibfnamefont {V.}~\bibnamefont {Giesz}}, \bibinfo {author} {\bibfnamefont {L.}~\bibnamefont {De~Santis}}, \bibinfo {author} {\bibfnamefont {J.~C.}\ \bibnamefont {Loredo}}, \bibinfo {author} {\bibfnamefont {M.~P.}\ \bibnamefont {Almeida}}, \bibinfo {author} {\bibfnamefont {G.}~\bibnamefont {Hornecker}}, \bibinfo {author} {\bibfnamefont {S.~L.}\ \bibnamefont {Portalupi}}, \bibinfo {author} {\bibfnamefont {T.}~\bibnamefont {Grange}}, \bibinfo {author} {\bibfnamefont {C.}~\bibnamefont {Antón}}, \bibinfo {author} {\bibfnamefont {J.}~\bibnamefont {Demory}}, \bibinfo {author} {\bibfnamefont {C.}~\bibnamefont {Gómez}}, \bibinfo {author} {\bibfnamefont {I.}~\bibnamefont {Sagnes}}, \bibinfo {author} {\bibfnamefont {N.~D.}\ \bibnamefont {Lanzillotti-Kimura}}, \bibinfo {author} {\bibfnamefont {A.}~\bibnamefont {Lemaítre}}, \bibinfo {author} {\bibfnamefont {A.}~\bibnamefont {Auffeves}}, \bibinfo {author}
  {\bibfnamefont {A.~G.}\ \bibnamefont {White}}, \bibinfo {author} {\bibfnamefont {L.}~\bibnamefont {Lanco}}, \ and\ \bibinfo {author} {\bibfnamefont {P.}~\bibnamefont {Senellart}},\ }\href {\doibase 10.1038/nphoton.2016.23} {\bibfield  {journal} {\bibinfo  {journal} {Nature Photonics}\ }\textbf {\bibinfo {volume} {10}},\ \bibinfo {pages} {340–345} (\bibinfo {year} {2016})}\BibitemShut {NoStop}%
\bibitem [{\citenamefont {Uppu}\ \emph {et~al.}(2020)\citenamefont {Uppu}, \citenamefont {Pedersen}, \citenamefont {Wang}, \citenamefont {Olesen}, \citenamefont {Papon}, \citenamefont {Zhou}, \citenamefont {Midolo}, \citenamefont {Scholz}, \citenamefont {Wieck}, \citenamefont {Ludwig},\ and\ \citenamefont {Lodahl}}]{Lohdal_QD}%
  \BibitemOpen
  \bibfield  {author} {\bibinfo {author} {\bibfnamefont {R.}~\bibnamefont {Uppu}}, \bibinfo {author} {\bibfnamefont {F.~T.}\ \bibnamefont {Pedersen}}, \bibinfo {author} {\bibfnamefont {Y.}~\bibnamefont {Wang}}, \bibinfo {author} {\bibfnamefont {C.~T.}\ \bibnamefont {Olesen}}, \bibinfo {author} {\bibfnamefont {C.}~\bibnamefont {Papon}}, \bibinfo {author} {\bibfnamefont {X.}~\bibnamefont {Zhou}}, \bibinfo {author} {\bibfnamefont {L.}~\bibnamefont {Midolo}}, \bibinfo {author} {\bibfnamefont {S.}~\bibnamefont {Scholz}}, \bibinfo {author} {\bibfnamefont {A.~D.}\ \bibnamefont {Wieck}}, \bibinfo {author} {\bibfnamefont {A.}~\bibnamefont {Ludwig}}, \ and\ \bibinfo {author} {\bibfnamefont {P.}~\bibnamefont {Lodahl}},\ }\href {\doibase 10.1126/sciadv.abc8268} {\bibfield  {journal} {\bibinfo  {journal} {Science Advances}\ }\textbf {\bibinfo {volume} {6}},\ \bibinfo {pages} {eabc8268} (\bibinfo {year} {2020})}\BibitemShut {NoStop}%
\bibitem [{\citenamefont {Cirac}\ \emph {et~al.}(1999)\citenamefont {Cirac}, \citenamefont {Ekert},\ and\ \citenamefont {Macchiavello}}]{Macchiavello_99}%
  \BibitemOpen
  \bibfield  {author} {\bibinfo {author} {\bibfnamefont {J.~I.}\ \bibnamefont {Cirac}}, \bibinfo {author} {\bibfnamefont {A.~K.}\ \bibnamefont {Ekert}}, \ and\ \bibinfo {author} {\bibfnamefont {C.}~\bibnamefont {Macchiavello}},\ }\href {\doibase 10.1103/PhysRevLett.82.4344} {\bibfield  {journal} {\bibinfo  {journal} {Physical Review Letters}\ }\textbf {\bibinfo {volume} {82}},\ \bibinfo {pages} {4344} (\bibinfo {year} {1999})}\BibitemShut {NoStop}%
\bibitem [{\citenamefont {Pan}\ \emph {et~al.}(2003)\citenamefont {Pan}, \citenamefont {Gasparoni}, \citenamefont {Ursin}, \citenamefont {Weihs},\ and\ \citenamefont {Zeilinger}}]{Pan2003}%
  \BibitemOpen
  \bibfield  {author} {\bibinfo {author} {\bibfnamefont {J.-W.}\ \bibnamefont {Pan}}, \bibinfo {author} {\bibfnamefont {S.}~\bibnamefont {Gasparoni}}, \bibinfo {author} {\bibfnamefont {R.}~\bibnamefont {Ursin}}, \bibinfo {author} {\bibfnamefont {G.}~\bibnamefont {Weihs}}, \ and\ \bibinfo {author} {\bibfnamefont {A.}~\bibnamefont {Zeilinger}},\ }\href {\doibase 10.1038/nature01623} {\bibfield  {journal} {\bibinfo  {journal} {Nature}\ }\textbf {\bibinfo {volume} {423}},\ \bibinfo {pages} {417} (\bibinfo {year} {2003})}\BibitemShut {NoStop}%
\bibitem [{\citenamefont {Ricci}\ \emph {et~al.}(2004)\citenamefont {Ricci}, \citenamefont {Martini}, \citenamefont {Cerf}, \citenamefont {Filip}, \citenamefont {Fiur\'a\ifmmode~\check{s}\else \v{s}\fi{}ek},\ and\ \citenamefont {Macchiavello}}]{DeMartini_04}%
  \BibitemOpen
  \bibfield  {author} {\bibinfo {author} {\bibfnamefont {M.}~\bibnamefont {Ricci}}, \bibinfo {author} {\bibfnamefont {F.~D.}\ \bibnamefont {Martini}}, \bibinfo {author} {\bibfnamefont {N.~J.}\ \bibnamefont {Cerf}}, \bibinfo {author} {\bibfnamefont {R.}~\bibnamefont {Filip}}, \bibinfo {author} {\bibfnamefont {J.}~\bibnamefont {Fiur\'a\ifmmode~\check{s}\else \v{s}\fi{}ek}}, \ and\ \bibinfo {author} {\bibfnamefont {C.}~\bibnamefont {Macchiavello}},\ }\href {\doibase 10.1103/PhysRevLett.93.170501} {\bibfield  {journal} {\bibinfo  {journal} {Physical Review Letters}\ }\textbf {\bibinfo {volume} {93}},\ \bibinfo {pages} {170501} (\bibinfo {year} {2004})}\BibitemShut {NoStop}%
\bibitem [{\citenamefont {Bravyi}\ and\ \citenamefont {Kitaev}(2005)}]{Kitaev_2005}%
  \BibitemOpen
  \bibfield  {author} {\bibinfo {author} {\bibfnamefont {S.}~\bibnamefont {Bravyi}}\ and\ \bibinfo {author} {\bibfnamefont {A.}~\bibnamefont {Kitaev}},\ }\href {\doibase 10.1103/PhysRevA.71.022316} {\bibfield  {journal} {\bibinfo  {journal} {Physical Review A}\ }\textbf {\bibinfo {volume} {71}},\ \bibinfo {pages} {022316} (\bibinfo {year} {2005})}\BibitemShut {NoStop}%
\bibitem [{\citenamefont {Salart}\ \emph {et~al.}(2010)\citenamefont {Salart}, \citenamefont {Landry}, \citenamefont {Sangouard}, \citenamefont {Gisin}, \citenamefont {Herrmann}, \citenamefont {Sanguinetti}, \citenamefont {Simon}, \citenamefont {Sohler}, \citenamefont {Thew}, \citenamefont {Thomas},\ and\ \citenamefont {Zbinden}}]{Salart_20210}%
  \BibitemOpen
  \bibfield  {author} {\bibinfo {author} {\bibfnamefont {D.}~\bibnamefont {Salart}}, \bibinfo {author} {\bibfnamefont {O.}~\bibnamefont {Landry}}, \bibinfo {author} {\bibfnamefont {N.}~\bibnamefont {Sangouard}}, \bibinfo {author} {\bibfnamefont {N.}~\bibnamefont {Gisin}}, \bibinfo {author} {\bibfnamefont {H.}~\bibnamefont {Herrmann}}, \bibinfo {author} {\bibfnamefont {B.}~\bibnamefont {Sanguinetti}}, \bibinfo {author} {\bibfnamefont {C.}~\bibnamefont {Simon}}, \bibinfo {author} {\bibfnamefont {W.}~\bibnamefont {Sohler}}, \bibinfo {author} {\bibfnamefont {R.~T.}\ \bibnamefont {Thew}}, \bibinfo {author} {\bibfnamefont {A.}~\bibnamefont {Thomas}}, \ and\ \bibinfo {author} {\bibfnamefont {H.}~\bibnamefont {Zbinden}},\ }\href {\doibase 10.1103/PhysRevLett.104.180504} {\bibfield  {journal} {\bibinfo  {journal} {Physical Review Letters}\ }\textbf {\bibinfo {volume} {104}},\ \bibinfo {pages} {180504} (\bibinfo {year} {2010})}\BibitemShut {NoStop}%
\bibitem [{\citenamefont {Sparrow}(2018)}]{sparrow2018quantum}%
  \BibitemOpen
  \bibfield  {author} {\bibinfo {author} {\bibfnamefont {C.}~\bibnamefont {Sparrow}},\ }\emph {\bibinfo {title} {Quantum interference in universal linear optical devices for quantum computation and simulation}},\ \href@noop {} {Ph.D. thesis},\ \bibinfo  {school} {Imperial College London} (\bibinfo {year} {2018})\BibitemShut {NoStop}%
\bibitem [{\citenamefont {Marshall}(2022)}]{Marshall_22}%
  \BibitemOpen
  \bibfield  {author} {\bibinfo {author} {\bibfnamefont {J.}~\bibnamefont {Marshall}},\ }\href {\doibase 10.1103/PhysRevLett.129.213601} {\bibfield  {journal} {\bibinfo  {journal} {Physical Review Letters}\ }\textbf {\bibinfo {volume} {129}},\ \bibinfo {pages} {213601} (\bibinfo {year} {2022})}\BibitemShut {NoStop}%
\bibitem [{\citenamefont {Somhorst}\ \emph {et~al.}(2025)\citenamefont {Somhorst}, \citenamefont {Sau\"er}, \citenamefont {van~den Hoven},\ and\ \citenamefont {Renema}}]{Somhorst25}%
  \BibitemOpen
  \bibfield  {author} {\bibinfo {author} {\bibfnamefont {F.}~\bibnamefont {Somhorst}}, \bibinfo {author} {\bibfnamefont {B.}~\bibnamefont {Sau\"er}}, \bibinfo {author} {\bibfnamefont {S.}~\bibnamefont {van~den Hoven}}, \ and\ \bibinfo {author} {\bibfnamefont {J.}~\bibnamefont {Renema}},\ }\href {\doibase 10.1103/PhysRevApplied.23.044003} {\bibfield  {journal} {\bibinfo  {journal} {Physical Review Applied}\ }\textbf {\bibinfo {volume} {23}},\ \bibinfo {pages} {044003} (\bibinfo {year} {2025})}\BibitemShut {NoStop}%
\bibitem [{\citenamefont {Saied}\ \emph {et~al.}(2025)\citenamefont {Saied}, \citenamefont {Marshall}, \citenamefont {Anand},\ and\ \citenamefont {Rieffel}}]{Saied25}%
  \BibitemOpen
  \bibfield  {author} {\bibinfo {author} {\bibfnamefont {J.}~\bibnamefont {Saied}}, \bibinfo {author} {\bibfnamefont {J.}~\bibnamefont {Marshall}}, \bibinfo {author} {\bibfnamefont {N.}~\bibnamefont {Anand}}, \ and\ \bibinfo {author} {\bibfnamefont {E.~G.}\ \bibnamefont {Rieffel}},\ }\href {\doibase 10.1103/PhysRevApplied.23.034079} {\bibfield  {journal} {\bibinfo  {journal} {Physical Review Applied}\ }\textbf {\bibinfo {volume} {23}},\ \bibinfo {pages} {034079} (\bibinfo {year} {2025})}\BibitemShut {NoStop}%
\bibitem [{\citenamefont {Faurby}\ \emph {et~al.}(2024)\citenamefont {Faurby}, \citenamefont {Carosini}, \citenamefont {Cao}, \citenamefont {Sund}, \citenamefont {Hansen}, \citenamefont {Giorgino}, \citenamefont {Villadsen}, \citenamefont {van~den Hoven}, \citenamefont {Lodahl}, \citenamefont {Paesani}, \citenamefont {Loredo},\ and\ \citenamefont {Walther}}]{Carosini_24}%
  \BibitemOpen
  \bibfield  {author} {\bibinfo {author} {\bibfnamefont {C.~F.~D.}\ \bibnamefont {Faurby}}, \bibinfo {author} {\bibfnamefont {L.}~\bibnamefont {Carosini}}, \bibinfo {author} {\bibfnamefont {H.}~\bibnamefont {Cao}}, \bibinfo {author} {\bibfnamefont {P.~I.}\ \bibnamefont {Sund}}, \bibinfo {author} {\bibfnamefont {L.~M.}\ \bibnamefont {Hansen}}, \bibinfo {author} {\bibfnamefont {F.}~\bibnamefont {Giorgino}}, \bibinfo {author} {\bibfnamefont {A.~B.}\ \bibnamefont {Villadsen}}, \bibinfo {author} {\bibfnamefont {S.~N.}\ \bibnamefont {van~den Hoven}}, \bibinfo {author} {\bibfnamefont {P.}~\bibnamefont {Lodahl}}, \bibinfo {author} {\bibfnamefont {S.}~\bibnamefont {Paesani}}, \bibinfo {author} {\bibfnamefont {J.~C.}\ \bibnamefont {Loredo}}, \ and\ \bibinfo {author} {\bibfnamefont {P.}~\bibnamefont {Walther}},\ }\href {\doibase 10.1103/PhysRevLett.133.033604} {\bibfield  {journal} {\bibinfo  {journal} {Physical Review Letters}\ }\textbf {\bibinfo {volume} {133}},\ \bibinfo {pages} {033604} (\bibinfo {year}
  {2024})}\BibitemShut {NoStop}%
\bibitem [{\citenamefont {Shchesnovich}\ and\ \citenamefont {Bezerra}(2018)}]{Schesnovic19}%
  \BibitemOpen
  \bibfield  {author} {\bibinfo {author} {\bibfnamefont {V.~S.}\ \bibnamefont {Shchesnovich}}\ and\ \bibinfo {author} {\bibfnamefont {M.~E.~O.}\ \bibnamefont {Bezerra}},\ }\href {\doibase 10.1103/PhysRevA.98.033805} {\bibfield  {journal} {\bibinfo  {journal} {Phys. Rev. A}\ }\textbf {\bibinfo {volume} {98}},\ \bibinfo {pages} {033805} (\bibinfo {year} {2018})}\BibitemShut {NoStop}%
\bibitem [{\citenamefont {Jones}\ \emph {et~al.}(2020)\citenamefont {Jones}, \citenamefont {Menssen}, \citenamefont {Chrzanowski}, \citenamefont {Wolterink}, \citenamefont {Shchesnovich},\ and\ \citenamefont {Walmsley}}]{Menssen_22}%
  \BibitemOpen
  \bibfield  {author} {\bibinfo {author} {\bibfnamefont {A.~E.}\ \bibnamefont {Jones}}, \bibinfo {author} {\bibfnamefont {A.~J.}\ \bibnamefont {Menssen}}, \bibinfo {author} {\bibfnamefont {H.~M.}\ \bibnamefont {Chrzanowski}}, \bibinfo {author} {\bibfnamefont {T.~A.~W.}\ \bibnamefont {Wolterink}}, \bibinfo {author} {\bibfnamefont {V.~S.}\ \bibnamefont {Shchesnovich}}, \ and\ \bibinfo {author} {\bibfnamefont {I.~A.}\ \bibnamefont {Walmsley}},\ }\href {\doibase 10.1103/PhysRevLett.125.123603} {\bibfield  {journal} {\bibinfo  {journal} {Physical Review Letters}\ }\textbf {\bibinfo {volume} {125}},\ \bibinfo {pages} {123603} (\bibinfo {year} {2020})}\BibitemShut {NoStop}%
\bibitem [{\citenamefont {Hong}\ \emph {et~al.}(1987)\citenamefont {Hong}, \citenamefont {Ou},\ and\ \citenamefont {Mandel}}]{Hong1987}%
  \BibitemOpen
  \bibfield  {author} {\bibinfo {author} {\bibfnamefont {C.~K.}\ \bibnamefont {Hong}}, \bibinfo {author} {\bibfnamefont {Z.~Y.}\ \bibnamefont {Ou}}, \ and\ \bibinfo {author} {\bibfnamefont {L.}~\bibnamefont {Mandel}},\ }\href {\doibase doi:10.1103/PhysRevLett.59.2044} {\bibfield  {journal} {\bibinfo  {journal} {Physical Review Letters}\ }\textbf {\bibinfo {volume} {59}},\ \bibinfo {pages} {2044} (\bibinfo {year} {1987})}\BibitemShut {NoStop}%
\bibitem [{\citenamefont {Oszmaniec}\ \emph {et~al.}(2024)\citenamefont {Oszmaniec}, \citenamefont {Brod},\ and\ \citenamefont {Galvão}}]{Oszmaniec2024}%
  \BibitemOpen
  \bibfield  {author} {\bibinfo {author} {\bibfnamefont {M.}~\bibnamefont {Oszmaniec}}, \bibinfo {author} {\bibfnamefont {D.~J.}\ \bibnamefont {Brod}}, \ and\ \bibinfo {author} {\bibfnamefont {E.~F.}\ \bibnamefont {Galvão}},\ }\href {\doibase 10.1088/1367-2630/ad1a27} {\bibfield  {journal} {\bibinfo  {journal} {New Journal of Physics}\ }\textbf {\bibinfo {volume} {26}},\ \bibinfo {pages} {013053} (\bibinfo {year} {2024})}\BibitemShut {NoStop}%
\bibitem [{\citenamefont {Rodari}\ \emph {et~al.}(2024)\citenamefont {Rodari}, \citenamefont {Fernandes}, \citenamefont {Caruccio}, \citenamefont {Suprano}, \citenamefont {Hoch}, \citenamefont {Giordani}, \citenamefont {Carvacho}, \citenamefont {Albiero}, \citenamefont {Giano}, \citenamefont {Corrielli}, \citenamefont {Ceccarelli}, \citenamefont {Osellame}, \citenamefont {Brod}, \citenamefont {Novo}, \citenamefont {Spagnolo}, \citenamefont {Galvão},\ and\ \citenamefont {Sciarrino}}]{rodari_24_counter}%
  \BibitemOpen
  \bibfield  {author} {\bibinfo {author} {\bibfnamefont {G.}~\bibnamefont {Rodari}}, \bibinfo {author} {\bibfnamefont {C.}~\bibnamefont {Fernandes}}, \bibinfo {author} {\bibfnamefont {E.}~\bibnamefont {Caruccio}}, \bibinfo {author} {\bibfnamefont {A.}~\bibnamefont {Suprano}}, \bibinfo {author} {\bibfnamefont {F.}~\bibnamefont {Hoch}}, \bibinfo {author} {\bibfnamefont {T.}~\bibnamefont {Giordani}}, \bibinfo {author} {\bibfnamefont {G.}~\bibnamefont {Carvacho}}, \bibinfo {author} {\bibfnamefont {R.}~\bibnamefont {Albiero}}, \bibinfo {author} {\bibfnamefont {N.~D.}\ \bibnamefont {Giano}}, \bibinfo {author} {\bibfnamefont {G.}~\bibnamefont {Corrielli}}, \bibinfo {author} {\bibfnamefont {F.}~\bibnamefont {Ceccarelli}}, \bibinfo {author} {\bibfnamefont {R.}~\bibnamefont {Osellame}}, \bibinfo {author} {\bibfnamefont {D.~J.}\ \bibnamefont {Brod}}, \bibinfo {author} {\bibfnamefont {L.}~\bibnamefont {Novo}}, \bibinfo {author} {\bibfnamefont {N.}~\bibnamefont {Spagnolo}}, \bibinfo {author} {\bibfnamefont {E.~F.}\
  \bibnamefont {Galvão}}, \ and\ \bibinfo {author} {\bibfnamefont {F.}~\bibnamefont {Sciarrino}},\ }\href {https://arxiv.org/abs/2410.15883} {\enquote {\bibinfo {title} {Experimental observation of counter-intuitive features of photonic bunching},}\ } (\bibinfo {year} {2024}),\ \Eprint {http://arxiv.org/abs/2410.15883} {arXiv:2410.15883 [quant-ph]} \BibitemShut {NoStop}%
\bibitem [{\citenamefont {Shchesnovich}(2015)}]{Shchesnovich2015}%
  \BibitemOpen
  \bibfield  {author} {\bibinfo {author} {\bibfnamefont {V.~S.}\ \bibnamefont {Shchesnovich}},\ }\href {\doibase 10.1103/physreva.91.013844} {\bibfield  {journal} {\bibinfo  {journal} {Physical Review A}\ }\textbf {\bibinfo {volume} {91}},\ \bibinfo {pages} {013844} (\bibinfo {year} {2015})}\BibitemShut {NoStop}%
\bibitem [{\citenamefont {Tichy}(2015)}]{Tichy2015}%
  \BibitemOpen
  \bibfield  {author} {\bibinfo {author} {\bibfnamefont {M.~C.}\ \bibnamefont {Tichy}},\ }\href {\doibase 10.1103/physreva.91.022316} {\bibfield  {journal} {\bibinfo  {journal} {Physical Review A}\ }\textbf {\bibinfo {volume} {91}},\ \bibinfo {pages} {022316} (\bibinfo {year} {2015})}\BibitemShut {NoStop}%
\bibitem [{\citenamefont {Levy}\ and\ \citenamefont {Shalit}(2014)}]{Levy2014}%
  \BibitemOpen
  \bibfield  {author} {\bibinfo {author} {\bibfnamefont {E.}~\bibnamefont {Levy}}\ and\ \bibinfo {author} {\bibfnamefont {O.~M.}\ \bibnamefont {Shalit}},\ }\href {\doibase 10.1216/rmj-2014-44-1-203} {\bibfield  {journal} {\bibinfo  {journal} {Rocky Mountain Journal of Mathematics}\ }\textbf {\bibinfo {volume} {44}},\ \bibinfo {pages} {203} (\bibinfo {year} {2014})}\BibitemShut {NoStop}%
\bibitem [{\citenamefont {Fernandes}\ \emph {et~al.}(2024)\citenamefont {Fernandes}, \citenamefont {Wagner}, \citenamefont {Novo},\ and\ \citenamefont {Galv\~ao}}]{Fernandes_Barg}%
  \BibitemOpen
  \bibfield  {author} {\bibinfo {author} {\bibfnamefont {C.}~\bibnamefont {Fernandes}}, \bibinfo {author} {\bibfnamefont {R.}~\bibnamefont {Wagner}}, \bibinfo {author} {\bibfnamefont {L.}~\bibnamefont {Novo}}, \ and\ \bibinfo {author} {\bibfnamefont {E.~F.}\ \bibnamefont {Galv\~ao}},\ }\href {\doibase 10.1103/PhysRevLett.133.190201} {\bibfield  {journal} {\bibinfo  {journal} {Physical Review Letters}\ }\textbf {\bibinfo {volume} {133}},\ \bibinfo {pages} {190201} (\bibinfo {year} {2024})}\BibitemShut {NoStop}%
\bibitem [{\citenamefont {Gazzano}\ \emph {et~al.}(2013)\citenamefont {Gazzano}, \citenamefont {Michaelis~de Vasconcellos}, \citenamefont {Arnold}, \citenamefont {Nowak}, \citenamefont {Galopin}, \citenamefont {Sagnes}, \citenamefont {Lanco}, \citenamefont {Lemaître},\ and\ \citenamefont {Senellart}}]{Gazzano2013}%
  \BibitemOpen
  \bibfield  {author} {\bibinfo {author} {\bibfnamefont {O.}~\bibnamefont {Gazzano}}, \bibinfo {author} {\bibfnamefont {S.}~\bibnamefont {Michaelis~de Vasconcellos}}, \bibinfo {author} {\bibfnamefont {C.}~\bibnamefont {Arnold}}, \bibinfo {author} {\bibfnamefont {A.}~\bibnamefont {Nowak}}, \bibinfo {author} {\bibfnamefont {E.}~\bibnamefont {Galopin}}, \bibinfo {author} {\bibfnamefont {I.}~\bibnamefont {Sagnes}}, \bibinfo {author} {\bibfnamefont {L.}~\bibnamefont {Lanco}}, \bibinfo {author} {\bibfnamefont {A.}~\bibnamefont {Lemaître}}, \ and\ \bibinfo {author} {\bibfnamefont {P.}~\bibnamefont {Senellart}},\ }\href {\doibase 10.1038/ncomms2434} {\bibfield  {journal} {\bibinfo  {journal} {Nature Communications}\ }\textbf {\bibinfo {volume} {4}},\ \bibinfo {pages} {1425} (\bibinfo {year} {2013})}\BibitemShut {NoStop}%
\bibitem [{\citenamefont {Thomas}\ \emph {et~al.}(2021)\citenamefont {Thomas}, \citenamefont {Billard}, \citenamefont {Coste}, \citenamefont {Wein}, \citenamefont {Priya}, \citenamefont {Ollivier}, \citenamefont {Krebs}, \citenamefont {Tazaïrt}, \citenamefont {Harouri}, \citenamefont {Lemaitre}, \citenamefont {Sagnes}, \citenamefont {Anton}, \citenamefont {Lanco}, \citenamefont {Somaschi}, \citenamefont {Loredo},\ and\ \citenamefont {Senellart}}]{Thomas2021}%
  \BibitemOpen
  \bibfield  {author} {\bibinfo {author} {\bibfnamefont {S.}~\bibnamefont {Thomas}}, \bibinfo {author} {\bibfnamefont {M.}~\bibnamefont {Billard}}, \bibinfo {author} {\bibfnamefont {N.}~\bibnamefont {Coste}}, \bibinfo {author} {\bibfnamefont {S.}~\bibnamefont {Wein}}, \bibinfo {author} {\bibnamefont {Priya}}, \bibinfo {author} {\bibfnamefont {H.}~\bibnamefont {Ollivier}}, \bibinfo {author} {\bibfnamefont {O.}~\bibnamefont {Krebs}}, \bibinfo {author} {\bibfnamefont {L.}~\bibnamefont {Tazaïrt}}, \bibinfo {author} {\bibfnamefont {A.}~\bibnamefont {Harouri}}, \bibinfo {author} {\bibfnamefont {A.}~\bibnamefont {Lemaitre}}, \bibinfo {author} {\bibfnamefont {I.}~\bibnamefont {Sagnes}}, \bibinfo {author} {\bibfnamefont {C.}~\bibnamefont {Anton}}, \bibinfo {author} {\bibfnamefont {L.}~\bibnamefont {Lanco}}, \bibinfo {author} {\bibfnamefont {N.}~\bibnamefont {Somaschi}}, \bibinfo {author} {\bibfnamefont {J.}~\bibnamefont {Loredo}}, \ and\ \bibinfo {author} {\bibfnamefont {P.}~\bibnamefont {Senellart}},\ }\href
  {\doibase 10.1103/physrevlett.126.233601} {\bibfield  {journal} {\bibinfo  {journal} {Physical Review Letters}\ }\textbf {\bibinfo {volume} {126}},\ \bibinfo {pages} {233601} (\bibinfo {year} {2021})}\BibitemShut {NoStop}%
\bibitem [{\citenamefont {Pentangelo}\ \emph {et~al.}(2024)\citenamefont {Pentangelo}, \citenamefont {Di~Giano}, \citenamefont {Piacentini}, \citenamefont {Arpe}, \citenamefont {Ceccarelli}, \citenamefont {Crespi},\ and\ \citenamefont {Osellame}}]{Pentangelo2024}%
  \BibitemOpen
  \bibfield  {author} {\bibinfo {author} {\bibfnamefont {C.}~\bibnamefont {Pentangelo}}, \bibinfo {author} {\bibfnamefont {N.}~\bibnamefont {Di~Giano}}, \bibinfo {author} {\bibfnamefont {S.}~\bibnamefont {Piacentini}}, \bibinfo {author} {\bibfnamefont {R.}~\bibnamefont {Arpe}}, \bibinfo {author} {\bibfnamefont {F.}~\bibnamefont {Ceccarelli}}, \bibinfo {author} {\bibfnamefont {A.}~\bibnamefont {Crespi}}, \ and\ \bibinfo {author} {\bibfnamefont {R.}~\bibnamefont {Osellame}},\ }\href {\doibase 10.1515/nanoph-2023-0636} {\bibfield  {journal} {\bibinfo  {journal} {Nanophotonics}\ }\textbf {\bibinfo {volume} {13}},\ \bibinfo {pages} {2259–2270} (\bibinfo {year} {2024})}\BibitemShut {NoStop}%
\end{thebibliography}

%

\end{document}